\documentclass[a4paper,fleqn,usenatbib]{mnras}

\usepackage{mathptmx}

\usepackage[T1]{fontenc}
\usepackage{ae,aecompl}

\usepackage{graphicx}	
\usepackage{amsmath}	
\usepackage{amssymb}	
\usepackage{textcomp}
\usepackage{color}
\usepackage{multicol}
\usepackage{longtable}
\usepackage{rotating}
\usepackage{pdflscape}
\usepackage{natbib}
\usepackage{enumitem}
\usepackage{appendix}
\usepackage{url}
\usepackage{float}
\usepackage{tabularx}
\usepackage{booktabs}
\usepackage[caption = false]{subfig}
\usepackage{scalerel}

\title[DESI DR1 Cosmic Chronometer view on H(z)]{Measuring the expansion history of the Universe with DESI Cosmic Chronometers}

\author[Loubser]{S. Ilani Loubser$^{1,2}$\thanks{E-mail:Ilani.Loubser@nwu.ac.za (SIL)} \\
$^{1}$Centre for Space Research, North-West University, Potchefstroom 2520, South Africa\\
$^{2}$National Institute for Theoretical and Computational Sciences (NITheCS), Potchefstroom 2520, South Africa\\
}

\date{Accepted 2025 November 04. Received 2025 October 21; in original form 2025 September 04}

\pubyear{2025}

\begin{document}
\label{firstpage}
\pagerange{\pageref{firstpage}--\pageref{lastpage}}
\maketitle

\begin{abstract}
Studying large samples of massive, passively evolving galaxies (called cosmic chronometers, CC) provides us with the unique ability to measure the Universe's expansion history without assuming a cosmological model. The Dark Energy Spectroscopic Instrument (DESI) DR1 is currently the largest, publicly available, homogeneous set of galaxies with reliable spectroscopic redshifts, and covers a wide range in redshift. We extracted all massive galaxies (stellar mass $\log M_{\star}/M_{\sun} > 10.75$, and velocity dispersion $\sigma > 280$ km s$^{-1}$), with no emission in [OII] $\lambda$3727 \AA{}, with reliable redshifts as well as reliable D4000$_{\rm n}$ measurements from DR1. From this sample of 360 000 massive, passive galaxies, we used D4000$_{\rm n}$ and the method of cosmic chronometers to get three new direct, independent measurements of $H(z)=$ 88.48 $\pm\ 0.57(\rm stat) \pm 12.32(\rm syst)$, $H(z)=$ 119.45 $\pm\ 6.39(\rm stat) \pm 16.64(\rm syst)$, and $H(z)= 108.28 \pm 10.07(\rm stat) \pm 15.08(\rm syst)$ $\rm km\ s^{-1}\ Mpc^{-1}$ at $z=0.46$,  $z=0.67$, and $z=0.83$, respectively. This sample, which covers $0.3 < z < 1.0$, is the largest CC sample to date, and we reach statistical uncertainties of 0.65\%, 5.35\%, and 9.30\% on our three measurements. Our measurements show no significant tension with the \textit{Planck} $\Lambda$CDM cosmology. In our analysis, we also illustrate that even amongst samples of massive, passive galaxies, the effect of downsizing can clearly be seen.
\end{abstract}


\begin{keywords}
galaxies: evolution, galaxies: statistics, galaxies: distances and redshifts, (cosmology:) cosmological parameters
\end{keywords}


\section{Introduction}
\label{introduction}

The $\Lambda$CDM cosmological model, featuring a cosmological constant ($\Lambda$) and cold dark matter (CDM), provides a consistent framework for understanding the Universe's evolution from its primordial beginnings to its large-scale structure today. Its parameters have been constrained with remarkable precision by observations of the Cosmic Microwave Background (CMB), most notably by the \textit{Planck} satellite \citep{Planck2018params}. However, this picture is now challenged by a series of persistent tensions between early-universe predictions and late-universe observations. The most well-known, the ``Hubble tension" describes a significant discrepancy ($\sim$4$\sigma$) between the value of the Hubble constant ($H_{0}$), the present-day expansion rate of the Universe, measured locally using Type Ia supernovae calibrated with Cepheids (late-universe) versus the value inferred from cosmic microwave background (early-universe) observations \citep{Verde2019, DiValentino2021}. The late-universe observations rely on the traditional cosmic distance ladder, whereas the early-universe observations assume a specific cosmological model (such as $\Lambda$CDM). While these tensions could be systematic, they have also ignited speculation about new physics beyond the standard model, such as evolving dark energy, or modifications to gravity.

Alternative cosmological probes, such as cosmic chronometers (CC), play an important role in obtaining additional independent measurements to assess the reliability of the current models. It has become evident that a single probe is not adequate to constrain the properties and evolution of the Universe accurately \citep{Jiao2023}. The method of CCs does not need to assume a specific cosmological model and is essentially a measurement of time, as it measures the age evolution of a passively evolving population of galaxies, rather than relying on distances \citep{Moresco2016}. That makes it a very novel and powerful probe that offers a direct and independent measurement of the Hubble parameter $H(z)$, which describes the expansion rate of the Universe as a function of redshift. This method, introduced by \citet{Jimenez2002}, relies only on the minimal assumption of a Friedmann-Lema\^{i}tre-Robertson-Walker (FLRW) metric, and measures the differential age evolution of passive galaxies over a small redshift interval, making the method more robust than absolute age measurements of galaxies. 

In the standard FLRW spacetime metric, the Hubble parameter, $H(z)$, is related to the differential ageing of the Universe ($dt_{U}$) via \begin{equation} \label{eqn:Hz} H(z) = - \frac{1}{1+z} \frac{dz}{dt_{U}}. \end{equation} Therefore, by measuring the age difference between two passively evolving galaxies that formed at the same time but are separated by a small redshift interval ($dz$), one can calculate the derivative $dz/dt_{U}$, which then determines the Hubble parameter at that redshift \citep{Stern2010, Moresco2012, Moresco2022}. There are different methods that can be employed to measure the differential age evolution, e.g., full-spectrum fitting \citep{Ratsimbazafy2017, Borghi2022a, Borghi2022b}, Lick indices \citep{Jiao2023}, or the D4000$_{\rm n}$ index \citep{Moresco2024, Loubser2025}. In \citet{Loubser2025}, we used D4000$_{\rm n}$ \citep{Balogh1999} instead of full-spectrum fitting to accurately separate the various contributions to the systematic uncertainties. When we measure spectral indices, we select a smaller, specific wavelength region in the spectrum, and that increases statistical errors on the individual measurements since less spectral information is used. However, it avoids some of the systematics related to assumptions made in full-spectrum fitting, and if very large samples can be used for the measurements, it compensates for the lower statistical error on individual direct index measurements. 

The best probes for these CC differential age measurements are the most massive, passive galaxies, where the vast majority of the stars formed very early and over short timescales, with no significant star formation since \citep{Renzini2006, Thomas2010}. So far, existing CC measurements have been hampered by limited large samples of massive, passive galaxies over a wide range of redshifts, but large spectroscopic surveys like the Dark Energy Spectroscopic Instrument (DESI, \citealt{DESIDR1}), and soon the 4-metre Multi-Object Spectroscopic Telescope (4MOST, \citealt{deJong2022}), are starting to provide unprecedented statistics. This now enables a significant boost in the accuracy of $H(z)$ measurements made from CC. To make a good CC measurement, we need to probe the oldest, most massive galaxies at every redshift. In our previous study \citep{Loubser2025}, we used 53 Brightest Cluster Galaxies (BCGs) to make a CC measurement at $z = 0.5$, where the aim was to choose the most massive galaxies at every redshift a priori. We showed that the use of BCGs can decrease the systematic uncertainties of the CC approach, but the statistical uncertainty of the measurement was 47\% due to the limited sample size. 
 
DESI DR1 now offers the largest sample of extragalactic redshifts ever assembled \citep{DESIDR1}, and it allows for the identification of massive, passive galaxies in unprecedented numbers and with high fidelity across a wide redshift range. Furthermore, the high signal-to-noise ratio of DESI spectra enables precise measurements of the D4000$_{\rm n}$ index and other galactic properties. In this paper, we now follow the opposite approach to choosing BCGs a priori. We study a very large sample of massive, passive galaxies (not only BCGs), which decreases the statistical uncertainty significantly but has a very slight increase in the systematic uncertainty. The paper is structured as follows: Section \ref{data} describes the data, sample selection, and overall sample properties. Section \ref{downsizing} illustrates the effect of downsizing, even amongst the most massive galaxies, and the measurement of the D4000$_{\rm n} - z$ relation. The calibration of the relation is shown in Section \ref{calibration}, and the sources of systematic uncertainties are included in Section \ref{errors}. Section \ref{comparisons} shows the comparison to previous CC measurements. We also place our new $H(z)$ measurements in a cosmological context, comparing them to models to illustrate the possible applications. Section \ref{deprojection} shows the deprojection to $H_{0}$, and we summarise our results in Section \ref{summary}. In addition, in Appendix \ref{apertures}, we use Integral Field Unit (IFU) data from the Sloan Digital Sky Survey-IV (SDSS) Mapping Nearby Galaxies at APO (MaNGA) survey \citep{Bundy2015} to determine whether aperture effects can potentially affect DESI D4000$_{\rm n}$ index measurements, and in Appendix \ref{broadening} we also consider the possible effect of broadening on the D4000$_{\rm n}$ index measurements. 

\section{Data}
\label{data}

\subsection{The Dark Energy Spectroscopic Instrument (DESI)} 

The DESI Collaboration is executing a 5-year spectroscopic redshift survey to create a detailed map of the evolving structure of the Universe between redshift $z$ = 0 and $z$ = 4 \citep{Adame2024EDR, DESIDR1}. We use DESI Data Release 1 (DR1, also referred to as ``Iron''), which consists of all data acquired during the first 13 months of the DESI main survey (MAIN sample), as well as a uniform reprocessing of the DESI Survey Validation data that were previously made public in the DESI Early Data Release (SV sample). The DR1 main survey alone includes high-confidence redshifts for 18.7 million objects, of which 13.1 million are spectroscopically classified as galaxies. 

DESI operates with 5000 robotic fibres and a field of view of 3.2 degrees in diameter, and the wavelength coverage spans 3600 -- 9824 \AA{} \citep{DESIDR1}. Among DESI's five classes of primary spectroscopic targets, the Luminous Red Galaxies (LRGs), targeted in the redshift range of approximately 0.4 < $z$ < 1.1, are particularly relevant for CC studies \citep{Zhou2023}. However, we go through all ``\texttt{GALAXY}" spectra observed in dark time, as described below. The redshifts and spectral classifications (``\texttt{STAR}'', ``\texttt{GALAXY}'', ``\texttt{QSO}'') are determined by the \texttt{Redrock}\footnote{\url{https://github.com/desihub/redrock}} software, which fits principal component analysis (PCA) templates \citep{Guy2023, Adame2024EDR}. 

DESI data releases include numerous Value-Added Catalogues (VACs) that are built on core data products (spectra, classifications, redshifts; \citealt{DESIDR1}). For CC studies, the VAC from \texttt{FastSpecFit}\footnote{\url{https://fastspecfit.readthedocs.io/en/latest/fastspec.html}} containing spectrophotometric fitting results, e.g. stellar velocity dispersions ($\sigma$), stellar masses ($M_{*}$), $K$ corrections, magnitudes, and emission line equivalent widths is highly valuable for using homogeneous measurements for age dating \citep{fastspecfit, Moustakas2023}. In particular, we use the \texttt{FastSpecFit v3.0} catalogs\footnote{\url{https://data.desi.lbl.gov/public/dr1/vac/dr1/fastspecfit/iron/v3.0/catalogs/}}. The paper containing the detailed description of the \texttt{FastSpecFit} VAC is in preparation (Moustakas et al., in prep).

\subsection{Sample selection and composition}
\label{sample}

\begin{figure*}
\centering
\includegraphics[scale=0.386]{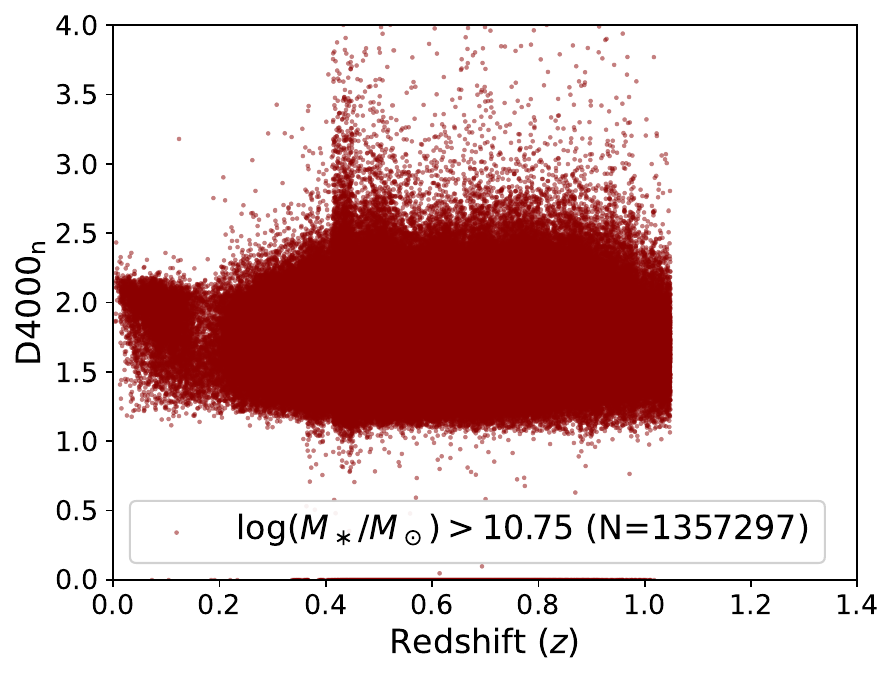}
\includegraphics[scale=0.386]{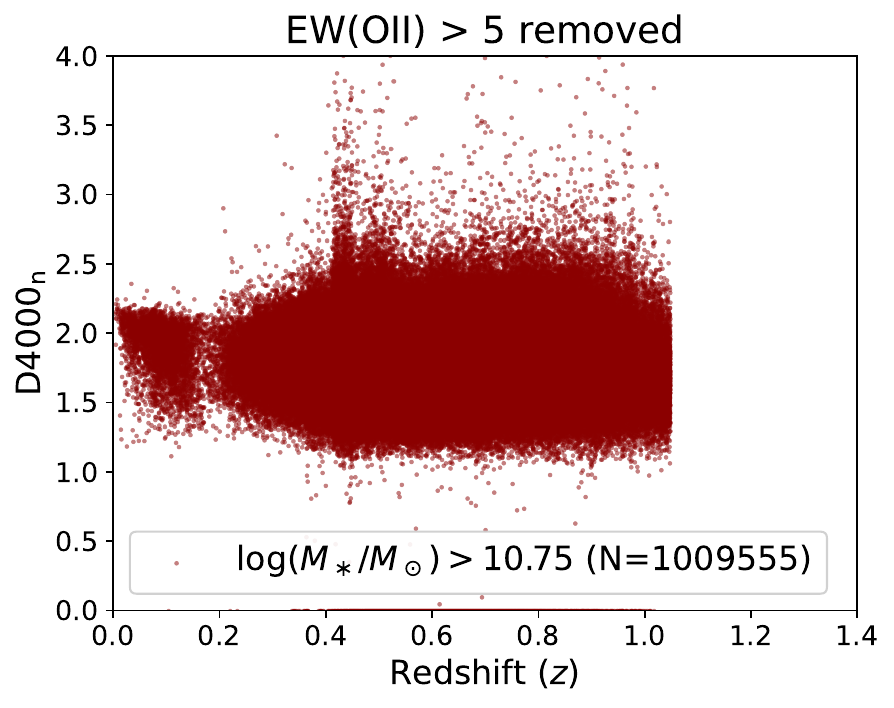}
\includegraphics[scale=0.386]{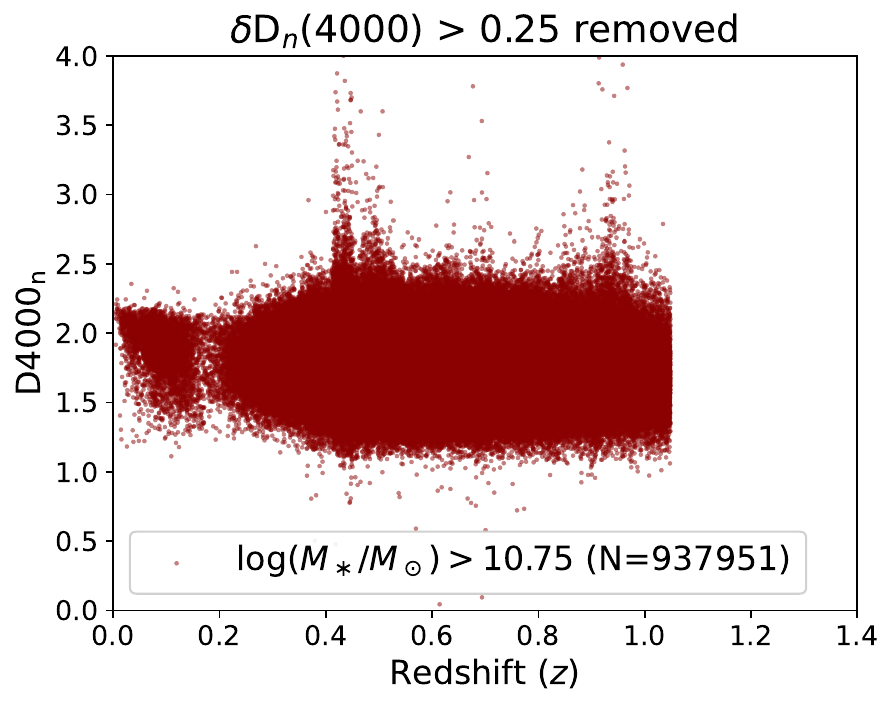}
   \caption{The sample selection, illustrated in the D4000$_{\rm n}$--redshift plane, as detailed in Section \ref{sample}. The number of galaxies retrieved with each selection cut is given in the legend.}
\label{fig:selection}
\end{figure*}

Rigorous selection of purely passively evolving galaxies is necessary to minimise contamination from younger star-forming systems and to ensure reliable differential age (or D4000$_{\rm n}$) measurements \citep{Moresco2018}. We therefore use a few steps and properties to select our main sample, before dividing it further into velocity dispersion bins, and examining the sample properties, in Section \ref{properties}:  

Step 1: By survey design, all LRGs are assigned to dark-time observations. We go through all ``dark time" data (the SV and MAIN samples that form part of DR1), and first select all objects classified as ``\texttt{GALAXY}". Another crucial quality flag is ``\texttt{ZWARN=0}", which indicates that there are no known problems with the input spectroscopic data or the redshift fit. This criterion is generally recommended for selecting objects with reliable redshifts \citep{Adame2024EDR, DESIDR1}.

Step 2: We then remove objects where the velocity dispersion could not be fitted from the spectra. These objects were assigned a nominal/default velocity dispersion of 250 km s$^{-1}$ (also see Appendix \ref{velocitydisp}). It is particularly the higher redshift galaxies for which velocity dispersions could not be accurately fitted, and the removal of these objects effectively results in a rather sharp cut-off of objects at approximately $z=1.05$ that can be seen in Figures \ref{fig:selection} and \ref{fig:selection2} (top panel). By survey design, the LRGs were also selected to be below $z<1.1$. 

Step 3: We also make a cut in stellar mass following previous CC studies, e.g., \citet{Moresco2016}, and select galaxies with stellar mass $\log M_{\star}/M_{\sun} > 10.75$ (see Figure \ref{fig:selection}, left panel). The stellar masses were derived from spectral energy distribution (SED) fitting by \texttt{FastSpecFit} and using a \citet{Chabrier2003} initial mass function (IMF). In general, galaxies above $\log M_{\star}/M_{\sun} > 10.5$ are expected to be slow rotators and passive (e.g., \citealt{Veale2017}). 

Step 4: We next select galaxies without detectable emission lines, where we use the [OII] $\lambda$3727 \AA{} line to make the cut since, given the wavelength range, it is measurable throughout our entire redshift range. The instrumental resolution ($\lambda/\Delta \lambda$) of DESI, which varies from 2000 in the blue arm to approximately 5000 in the near-infrared arm, is sufficient to resolve the [OII] doublet. Similarly to \citet{Moresco2016, Tomasetti2023}, we use a cut-off in equivalent width of five, but we make a strict selection of only galaxies with an equivalent width EW([OII] $\lambda$3726) < 5 and EW([OII] $\lambda$3729) < 5 to select passive galaxies. This [OII] emission line is, like H$\alpha$, extremely sensitive to the presence of the youngest and hottest stars with ages < 10 Myr \citep{Loubser2024}, and this selection removes all galaxies with very recent star formation (Figure \ref{fig:selection}, middle panel). 

Step 5: We further select galaxies with accurate D4000$_{\rm n}$ measurements by excluding galaxies with an error on D4000$_{\rm n}$ > 0.25. This selects galaxies with a signal-to-noise ratio around D4000$_{\rm n}$ greater than 10 \citep{Moresco2016}. This is shown in the right panel of Figure \ref{fig:selection}. 

\subsection{Sample properties}
\label{properties}

\begin{figure}
\centering
\includegraphics[scale=0.53]{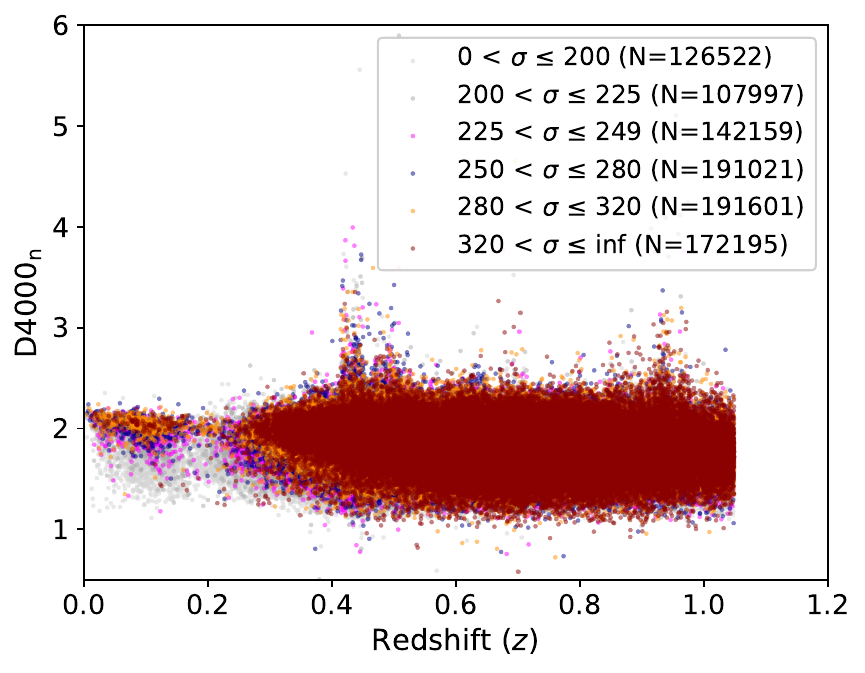} \\
\includegraphics[scale=0.35, clip]{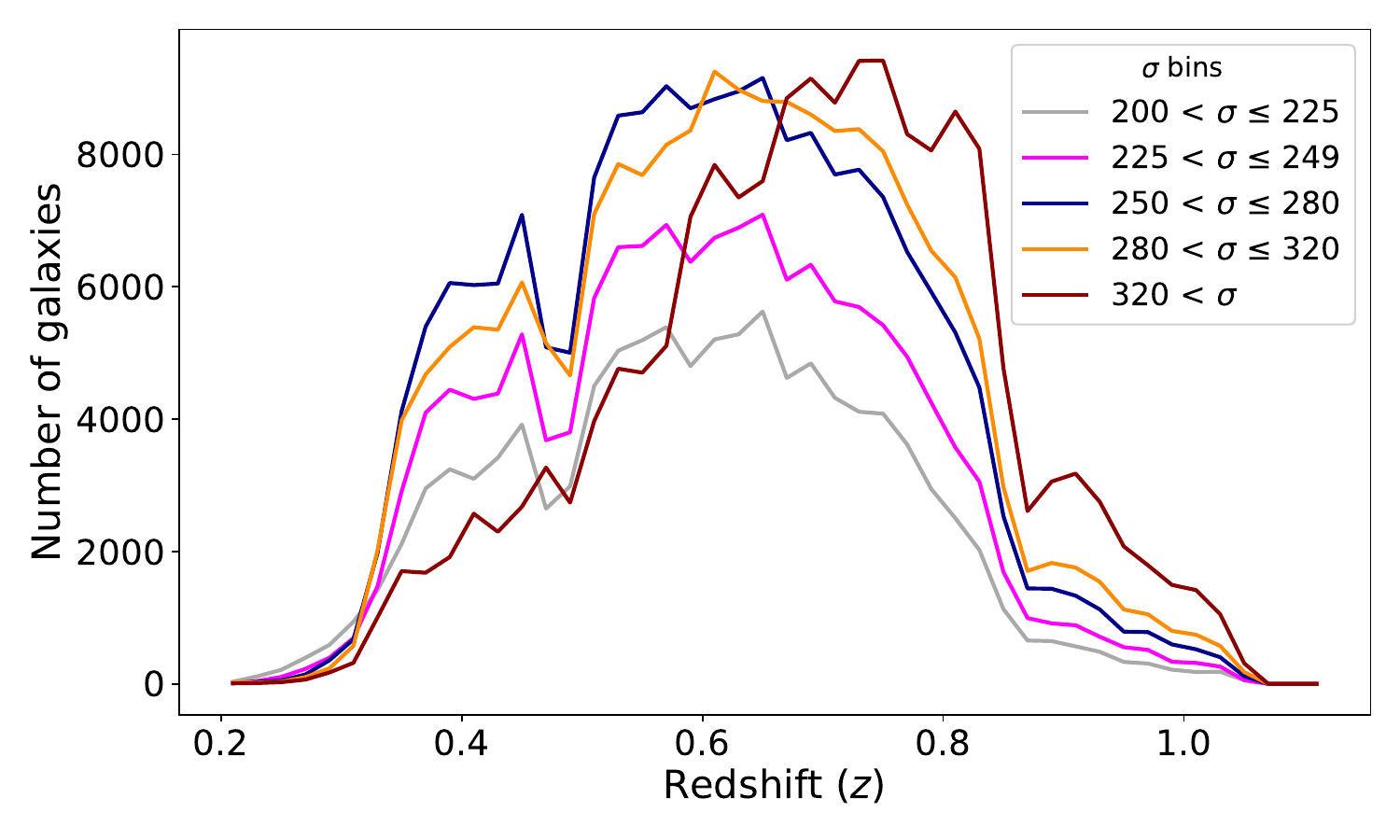}
   \caption{Top panel: The selected sample divided into velocity dispersion ($\sigma$) bins (with objects with the default velocity dispersion of 250 km s$^{-1}$ removed as shown in Appendix \ref{velocitydisp}). Lower panel: We bin redshift (from $z=$ 0.2 -- 1.1 in steps of 0.02), and we only use galaxies with $\sigma >$ 200 km s$^{-1}$. The plots illustrate why we fit the D4000$_{\rm n}$--$z$ relation (Section \ref{downsizing}) in the range $0.3 < z <1.0$.}
\label{fig:selection2}
\end{figure}

To further investigate the properties of galaxies selected through this sample selection, we divide the sample into velocity dispersion ($\sigma$) bins in Figure \ref{fig:selection2}, where we are particularly interested in the most massive galaxies, i.e., all galaxies with $\sigma > 280$ km s$^{-1}$ (the two highest velocity dispersion bins, consisting of 363 796 galaxies). These most massive galaxies are mostly distributed between $0.3 < z < 1.0$ (Figure \ref{fig:selection2}, bottom panel), with a sharp cut-off visible at $z \sim 1.05$ as discussed in Section \ref{sample}. 

\begin{figure}
\centering
\includegraphics[scale=0.362]{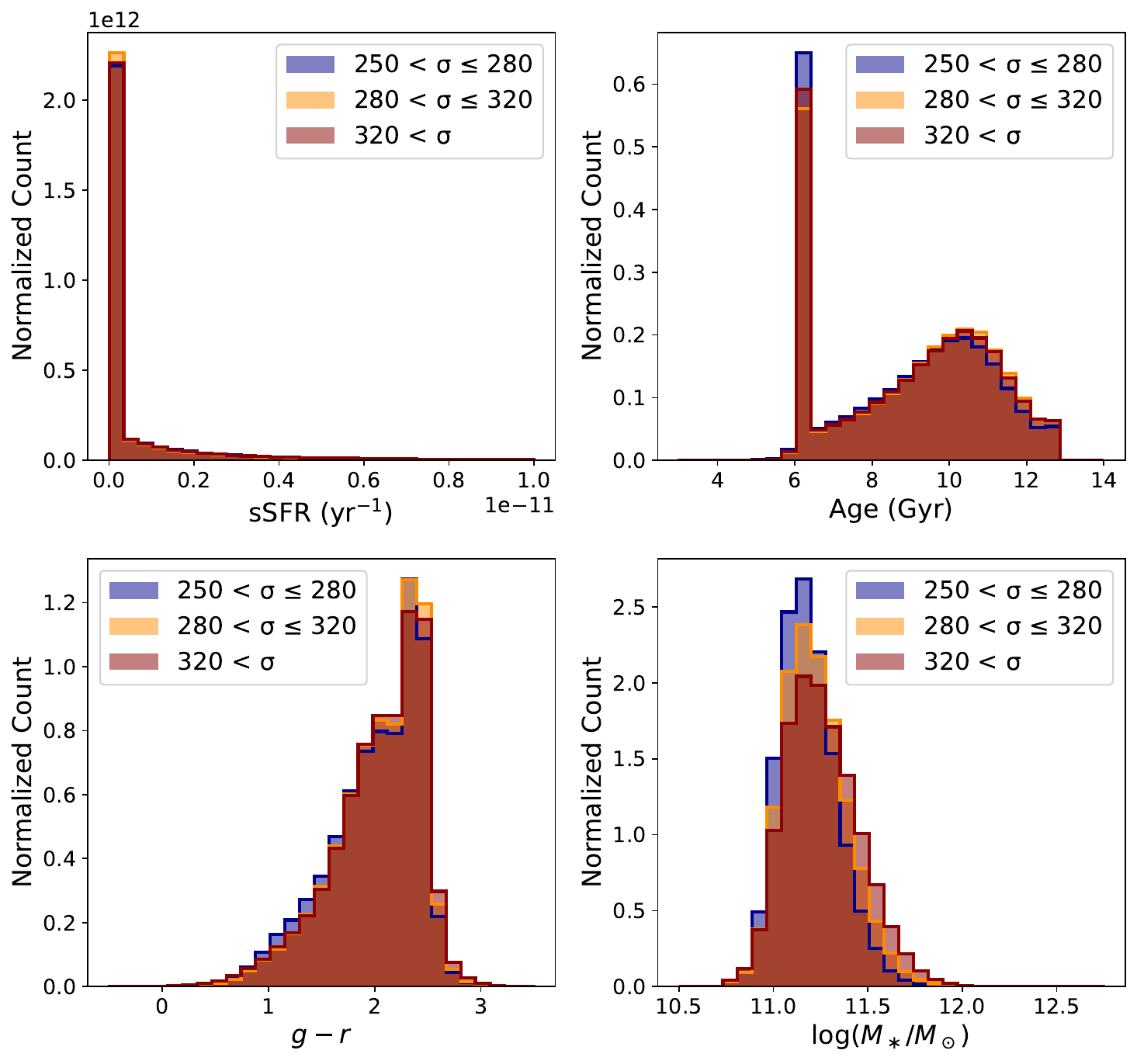}
   \caption{Distributions of sSFR, light-weighted SSP-equivalent Age, $g-r$ colours, and stellar mass ($\log M_{\star}/M_{\sun}$), for the massive galaxies ($\sigma$ > 250 km s$^{-1}$). The distributions for all three velocity dispersion bins are very similar.}
\label{fig:histograms}
\end{figure}

We prefer to use only spectroscopic properties for sample selection and not colours because of the effect of dust extinction, contamination, incompleteness, as well as the large redshift range and associated uncertain magnitude and colour corrections. Instead, we investigate the distributions of properties for the highest-mass galaxies to test our selection in Figure \ref{fig:histograms}, and to ensure that our selected sample is physically reasonable. Here, we show the three highest velocity dispersion bins, i.e., all objects with > 250 km s$^{-1}$, and look at their specific star formation rate (sSFR), their light-weighted simple stellar population equivalent (SSP) age, and their $g-r$ colours. We also include the distributions of stellar mass ($\log M_{\star}/M_{\sun}$). 

Figure \ref{fig:histograms} (top left panel) shows that sSFR is below $10^{-11}$ yr$^{-1}$, which is where galaxies are generally considered passive \citep{Wetzel2012, Furlong2015}. This also confirms the absence of objects with very recent or current star formation. To investigate the possible contamination by galaxies with star formation in the last $\sim$100 Myr to 1 Gyr, we consider the light-weighted simple stellar population-equivalent (SSP) Age and $g-r$ colour (Figure \ref{fig:histograms}, top right panel, and bottom left panel). The light-weighted SSP-equivalent age will be heavily biased towards younger (luminous) stars if any stellar populations < 1 Gyr are present \citep{Loubser2016}, and we find a negligible number of galaxies below an SSP-equivalent age of 5 Gyr. However, Figure \ref{fig:histograms} also shows that a significant fraction of the galaxies (in each of the bins) have SSP equivalent ages of $\sim$ 6 Gyr. The feature is artificial and stems from the fact that there is typically not enough information in the continuum spectra to derive accurate ages. This has no effect on our results as we do not use these fitted ages directly, instead we calibrate the D4000$_{\rm n}$ -- redshift relation using stellar population models themselves (Section \ref{calibration}). The templates used in \texttt{FastSpecFit} is the theoretical spectral library \texttt{C3K} available in the Flexible Stellar Population Synthesis (FSPS) models \citep{Conroy2009, Conroy2010}. If we disregard the peak, then we find a reasonably tight correlation between age and redshift, given the constraints, as expected. 

Fortunately, similar information about possible star formation in the last $\sim$100 Myr to 1 Gyr is contained in the $g-r$ colour (Figure \ref{fig:histograms}, bottom left panel). The colour $g-r$ is derived from the absolute magnitude in the DECam $g$ band and $r$ band from Legacy Imaging Surveys \citep{Dey2019}, band shifted to $z=1.0$ assuming $h=1.0$, by \texttt{FastSpecFit}. Lower values also indicate stellar populations younger than 1 Gyr \citep{Loubser2024}. The colours $g-r$ are in agreement with the colour distribution for the selection of the LRG sample shown in Figure 3 of \citet{Zhou2023}. Since the colours were corrected to $z=1.0$, we expect the distribution for passive, massive galaxies to have a peak > 1.7 mag (depending on the SED model assumed for the correction). In particular, the strong D4000$_{\rm n}$ is redward of both $g$ and $r$ at $z=1.0$, so both bands probe the steep UV slope of an old population. 

\subsection{D4000$_{\rm n}$ measurements}
\label{D4000measurements}

There are two effects that we need to consider that can potentially affect our D4000$_{\rm n}$ measurements. The first is that the 1.5$\arcsec$ diameter fibres of DESI cover the entire, or a large part, of galaxies at $z=1.0$, compared to covering a smaller part of massive galaxies at $z=0.3$. If there is a strong D4000$_{\rm n}$ gradient across the galaxy, there can be a D4000$_{\rm n}$ -- $z$ relation purely as a result of geometry. In Appendix \ref{apertures}, we use the IFU data of massive, passive galaxies in MaNGA to simulate DESI-sized apertures at different redshifts. The second effect is broadening, where an index measurement for a higher velocity dispersion galaxy can be broadened because of the velocity dispersion. We use SSP models to derive the broadening function for D4000$_{\rm n}$ in the Appendix \ref{broadening}. We show that both of these effects have a negligible influence on the results that follow.  

\section{Results}
\subsection{Downsizing and the D4000$_{\rm n} - z$ relation}
\label{downsizing}

\begin{figure*}
\centering
\includegraphics[scale=0.2665]{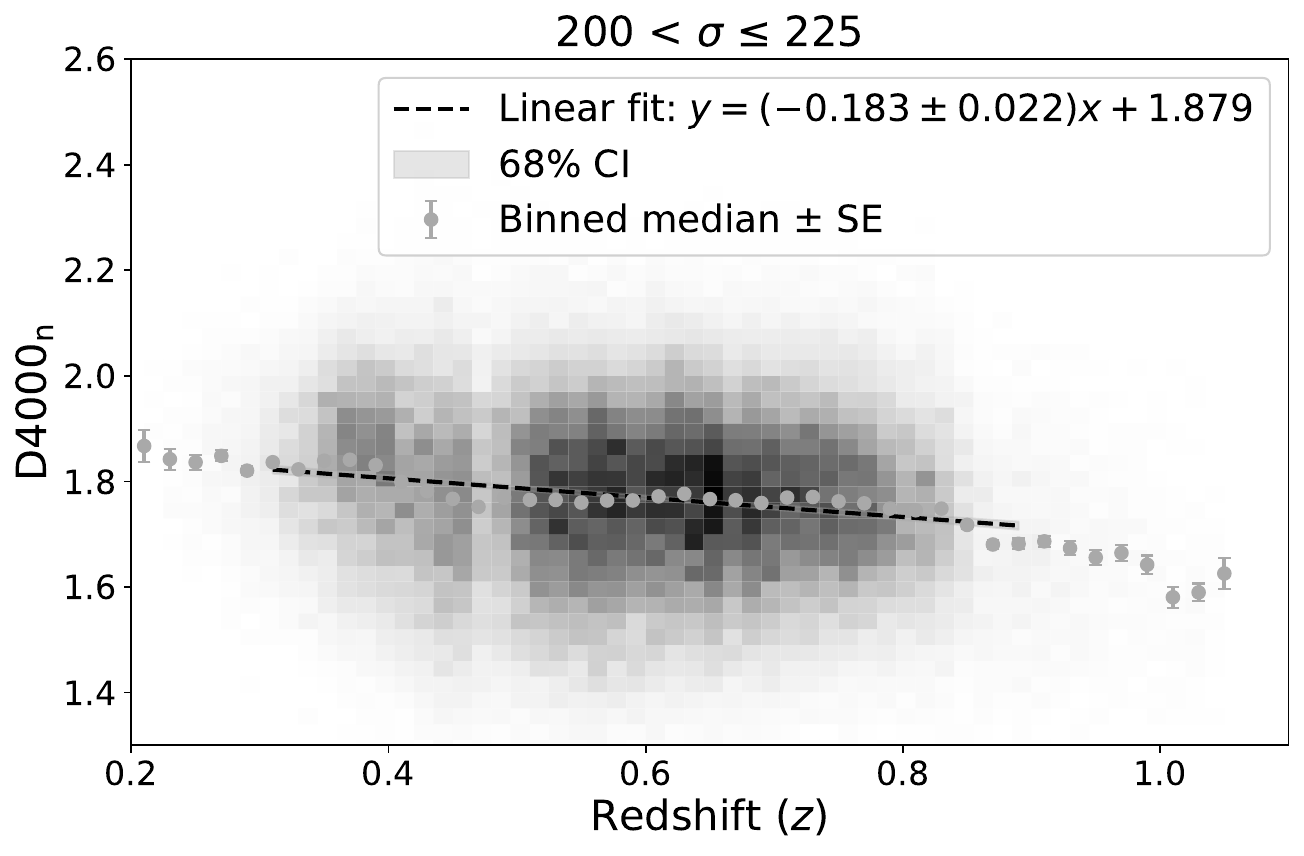}
\includegraphics[scale=0.2665]{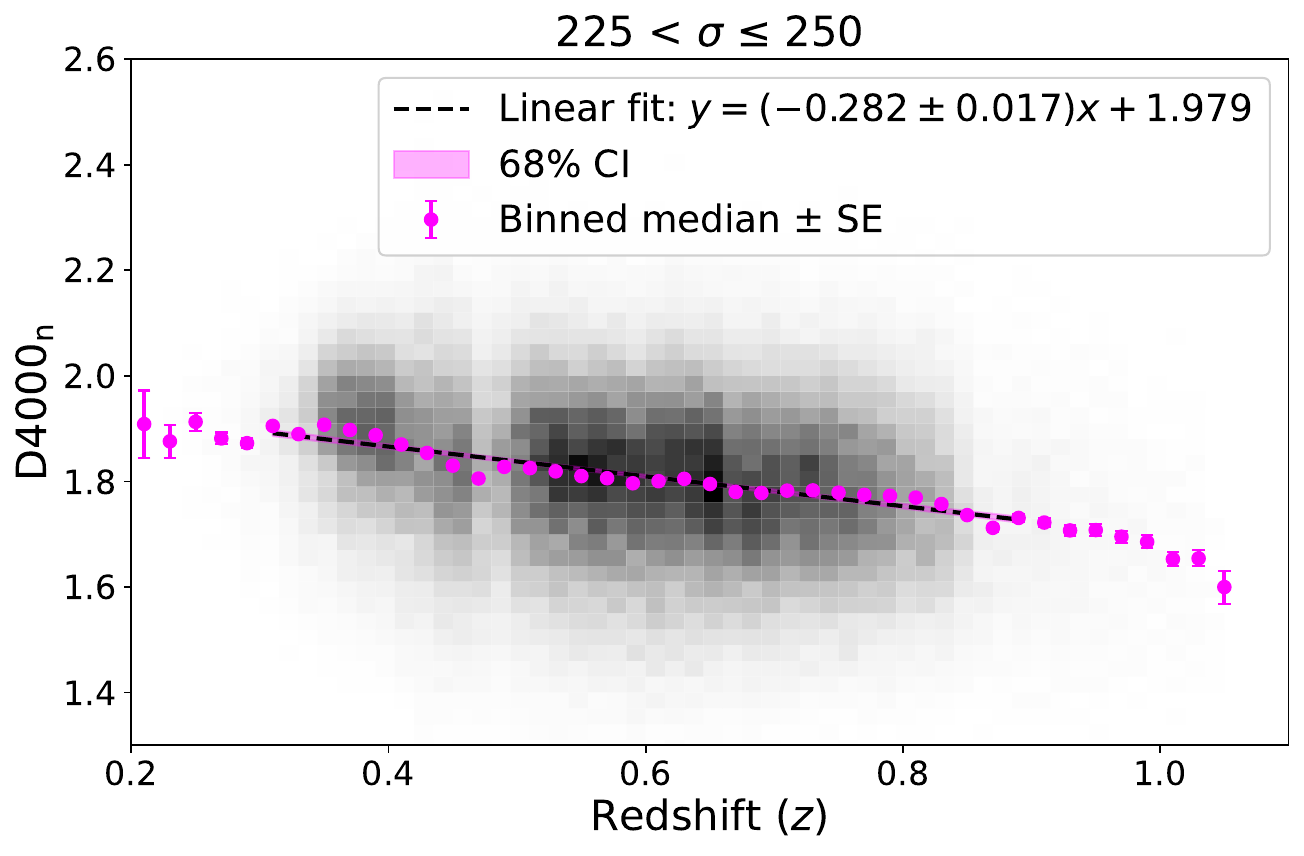}
\includegraphics[scale=0.2665]{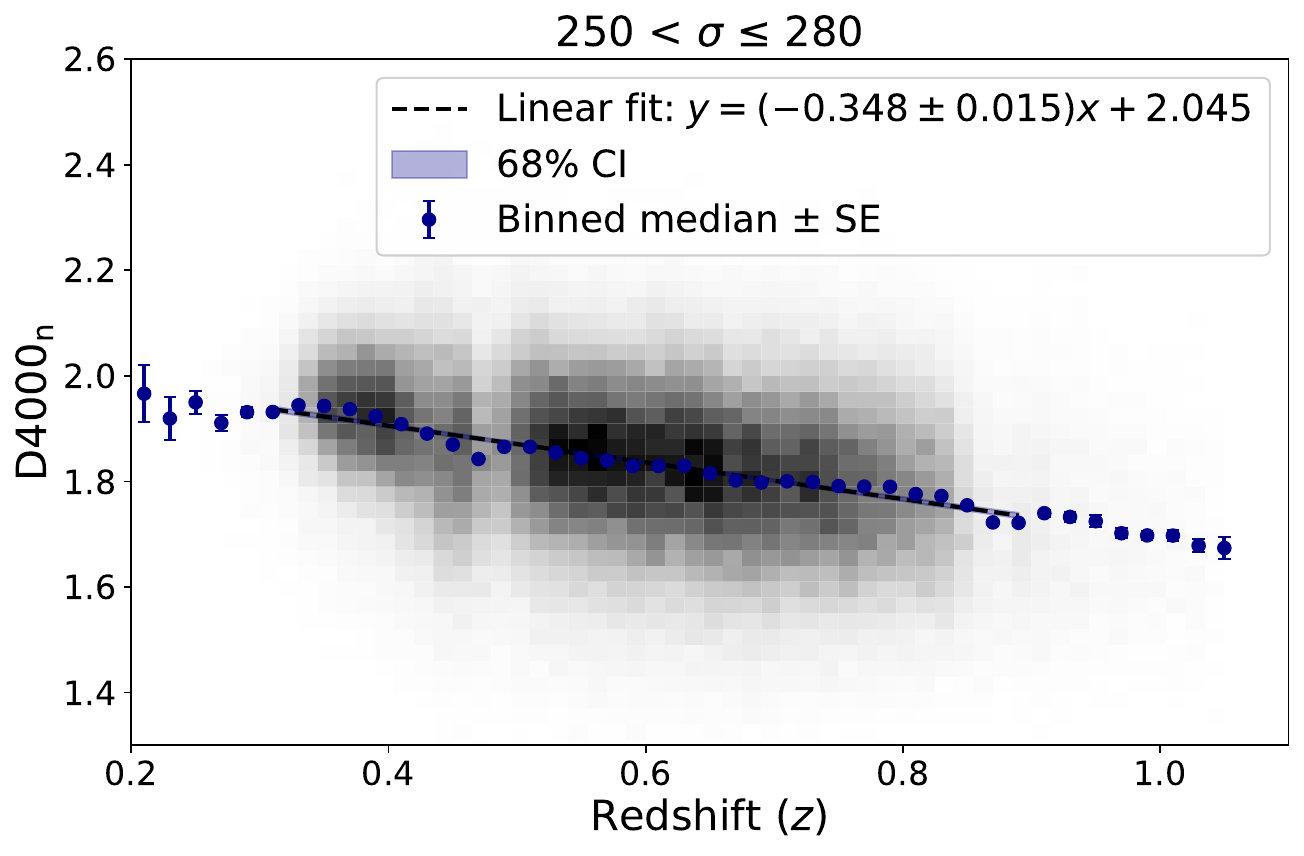} \\
\includegraphics[scale=0.2665]{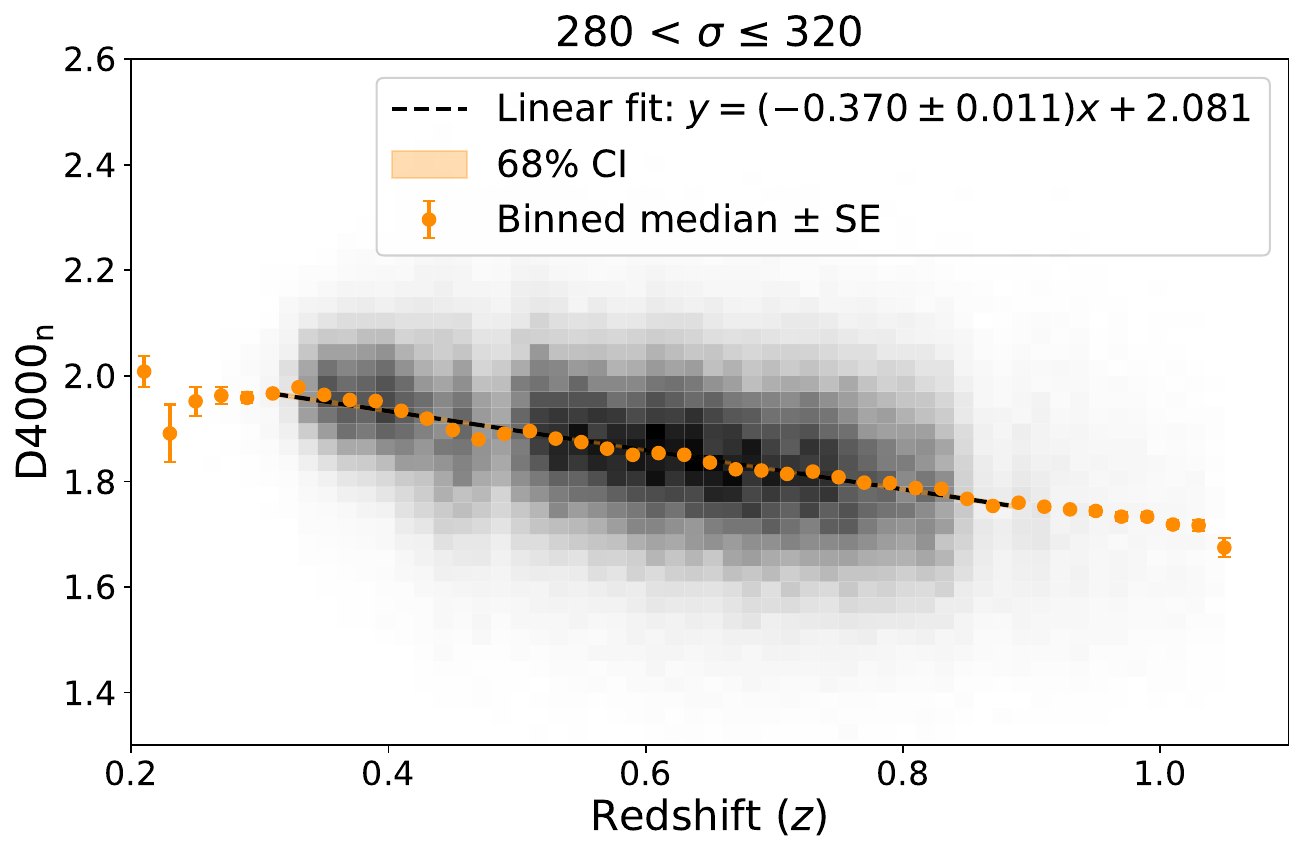}
\includegraphics[scale=0.2665]{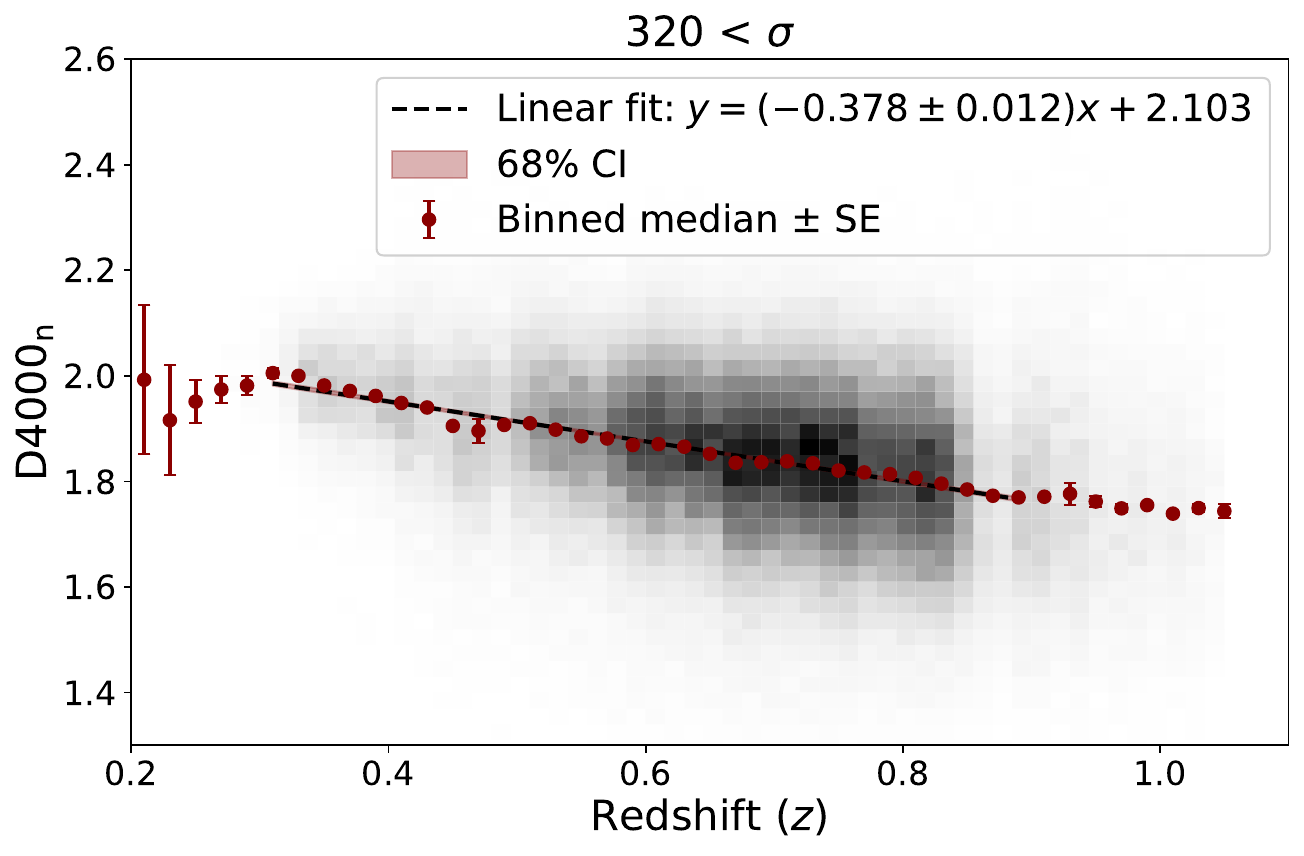}
\includegraphics[scale=0.2665]{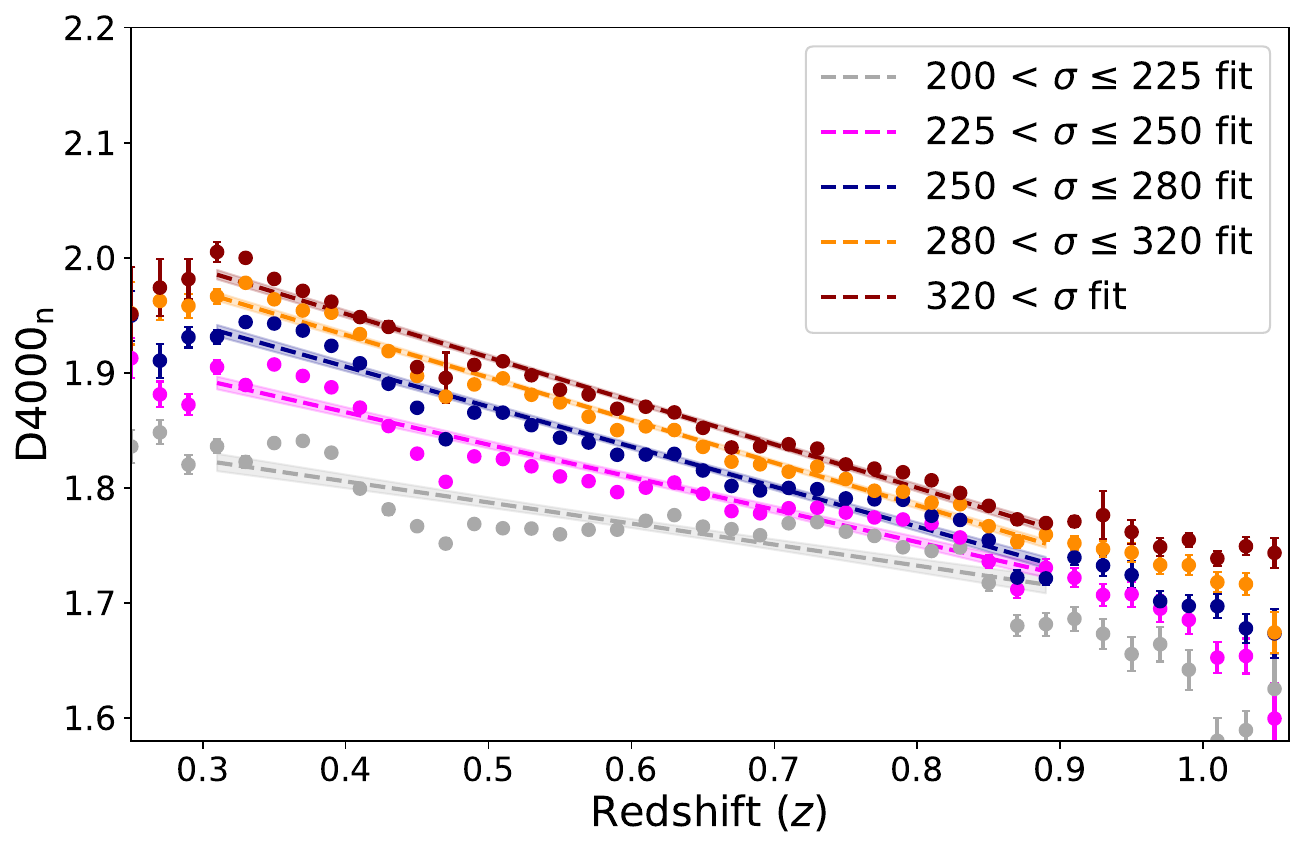}
   \caption{We divide the galaxies into velocity dispersion bins, and we bin the galaxies in redshift bins between $0.2 < z < 1.1$ where each bin is $\Delta z=0.02$ wide -- the density of galaxies in these bins are shown in grey-scale in the background. The straight lines fitted to the medians of the bins between $0.3 < z < 0.9$ (the range where there is a sufficient number of galaxies in all five bins) is plotted together in the bottom right panel, and clearly illustrates the effect of downsizing. The 68\% confidence levels on the fitted lines are shown in shaded bands.}
\label{fig:downsizing}
\end{figure*}

Based on Figure \ref{fig:selection2}, we divide the selected sample into five velocity dispersion bins (200 -- 225, 225 -- 250, 250 -- 280, 280 -- 320, and > 320 km s$^{-1}$), and we exclude all objects with the nominal velocity dispersion of 250 km s$^{-1}$. We bin the galaxies into redshift bins between $0.2 < z < 1.1$ where each bin is $\Delta z=0.02$ wide. The redshift bin width of $\Delta z=0.02$ is the maximum that should be used \citep{Borghi2022b, Loubser2025}, as it corresponds to a cosmic time at $z=0.6$ for which the D4000$_{\rm n}$ index is expected to change negligibly (the exact value depends on the stellar population model, metallicity, etc. as discussed in Section \ref{calibration}). For each redshift bin, we derive the median as well as the standard error on the median (SE), and use this to fit a single straight line between $0.3 < z < 0.9$ (the range where there is a sufficient number of galaxies in all five bins) for each velocity dispersion bin, as shown in Figure \ref{fig:downsizing}. The 68\% confidence levels on the fitted lines are shown with shaded bands, and the density of galaxies in the bins is shown in grey-scale in the background.

We then show the five fitted lines together in the last panel of Figure \ref{fig:downsizing} for easy comparison. It can be seen here that the slopes become shallower with decreasing velocity dispersion, and the data points more scattered. The figure illustrates how more massive galaxies are older, even among massive, passive galaxies. Galaxy mass drives galaxy evolution, with more massive galaxies forming their stars at earlier cosmic epochs compared to less massive ones \citep{Cowie1996, Thomas2010}. We therefore expect multiple age (or D4000$_{\rm n}$)--$z$ relations for different masses of populations, forming an envelope in the D4000$_{\rm n}$--$z$ plane. Figure \ref{fig:downsizing} also illustrates that this ‘downsizing’ effect is pronounced, as the difference between the slope of the bin with velocity dispersion > 320 km s$^{-1}$, and the 200 -- 225 km s$^{-1}$ bin has an 8$\sigma$ significance. This strong downsizing evidence justifies mass-dependent slopes and a strict velocity dispersion-threshold.

The statistical power of the DESI DR1 data allows us to quantitatively test the effect of relaxing the velocity dispersion threshold. We test a range of velocity dispersion thresholds, where we aimed to balance the statistical uncertainty on the slope and the systematic uncertainty that will be introduced by lowering the threshold. From Figure \ref{fig:downsizing} it can be seen that the two slopes for the 280 -- 320, and > 320 km s$^{-1}$ velocity dispersion bins are nearly identical, and combined the two bins contain more than 360 000 galaxies. We find that using the velocity dispersion $\sigma > $ 280 km s$^{-1}$ selection provides the optimal combination of precision and accuracy. We test this absence of mass dependence of the D4000$_{\rm n}$--$z$ slope when $\sigma > $ 280 km s$^{-1}$ is used, and confirm that the median values of D4000$_{\rm n}$ (binned in small velocity dispersion bins) show no correlation with velocity dispersion above this threshold. Figure \ref{fig:downsizing} shows that unless one has a very large sample of galaxies and reliable velocity dispersion measurements, there is a danger of mixing different galaxy mass populations which would directly have a significant (smearing) effect on the derived value of $H(z)$. The figure also illustrates that only the highest velocity dispersion bins (e.g., $\sigma$ > 280 km s$^{-1}$) should be used for accurate CC studies.

\begin{figure*}
\centering
\includegraphics[scale=0.55]{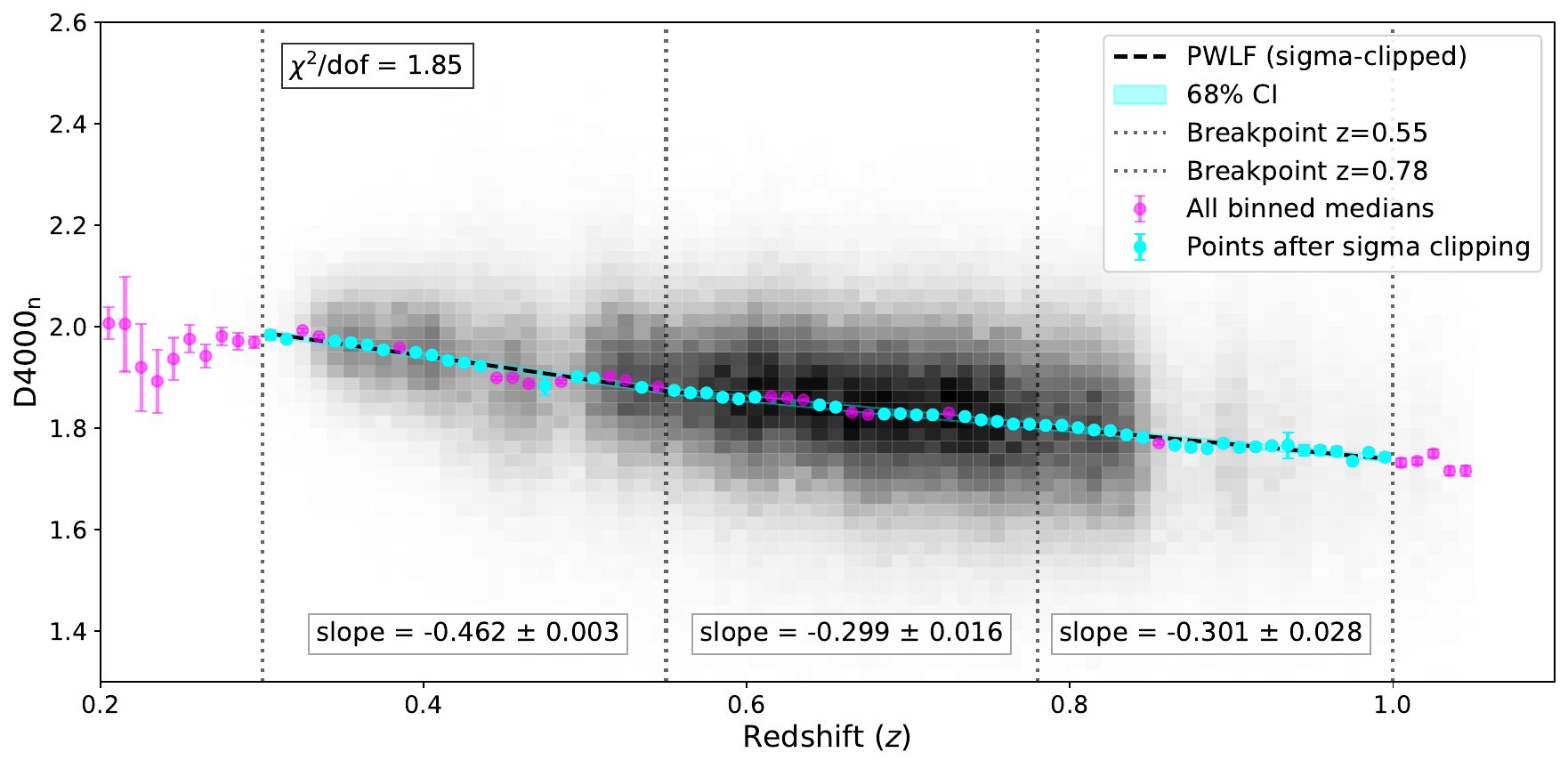}
   \caption{The D4000$_{\rm n} - z$ relation for galaxies with $\sigma > $ 280 km s$^{-1}$, with a redshift binning of $\Delta z = 0.01$. We fit a piece-wise linear function (PWLF) between $0.3 < z < 1.0$ to the median (and standard error on the median) values, using iterative sigma-slipping, with the breakpoints at $z=0.55$ and $z=0.78$.}
\label{fig:relation}
\end{figure*}

We test the sensitivity of the slope ($dD4000_{\rm n}/dz$) to the redshift binning, and find that $\Delta z = 0.01$ is optimal (best reduced $\chi^{2}_{\rm red}$), although the slope is not very sensitive to the binning used, as long as the maximum redshift binning of $\Delta z=0.02$ is not exceeded. We, therefore, use the median D4000$_{\rm n}$ (and the standard error on the median) in small redshift bins ($\Delta z = 0.01$), and because $H(z)$ evolves with redshift, we fit the relation in the range $0.3 < z < 1.0$ with a piecewise linear function adopting the public \texttt{PYTHON} code \texttt{pwlf}\footnote{\url{https://pypi.org/project/pwlf/}} by \citet{pwlf}.

To avoid residual contamination, we use iterative sigma-slipping to fit of the D4000$_{\rm n}$--$z$ relation. We use a sigma ($\sigma_{\rm c}$) of three for the clipping, and the fit converges within two iterations. We use the reduced $\chi^{2}_{\rm red}$ to test the optimum breakpoints, and their locations, for the piecewise linear function fitting. We balance between measuring as many $H(z)$ measurements as possible, over a range of redshifts, while keeping the $\chi^{2}_{\rm red}$ (and the statistical uncertainty on each of the slopes) low. We find that using two breakpoints (at $z=0.55$ and $z=0.78$) provides the best balance, with each of the three bins still having a minimum of 75 000 galaxies. We find that from the 70 data points (from $z=0.3$ to $z=1.0$ with $\Delta z = 0.01$), 17 are clipped with the 3$\sigma_{\rm c}$ threshold, resulting in a reduced $\chi^{2}_{\rm red}$ of 1.85 (compared to a reduced $\chi^{2}_{\rm red}$ of 6.3 when all 70 data points are used as a baseline).

This sigma-clipped fit between $0.3 < z < 1.0$, for galaxies with $\sigma > $ 280 km s$^{-1}$, is shown in Figure \ref{fig:relation} with the breakpoints at $z=0.55$ and $z=0.78$ indicated on the figure. The binned median data points clipped in the iterative sigma-slipping, as well as those that fall outside of the redshift range fitted, are shown in magenta, while the binned medians used in the fitting is shown in cyan. These are the three slopes ($dD4000_{\rm n}/dz$) that we use to derive the three $H(z)$ values. We take the median redshift for each of the three redshift bins rather than the bin centre, because the redshift distribution is skewed and non-uniform as shown in Figure \ref{fig:selection2} (bottom panel). The median redshift for 0.3 -- 0.55 bin is $z = 0.46$ (number of galaxies 93 198), and for the 0.55 -- 0.78 bin is $z = 0.67$ (number of galaxies 187 853), and for the 0.78 -- 1.00 bin is $z = 0.83$ (number of galaxies 77 208). Thus the total number of galaxies included in the fitting is 358 259, and the remaining 5 537 galaxies with $\sigma > $ 280 km s$^{-1}$ has redshifts outside our range of $0.3 < z < 1.0$. Our statistical uncertainties from the fit in Figure \ref{fig:relation} is 0.65\% on the slope ($dD4000_{\rm n}/dz$) at $z = 0.46$, 5.35\% at $z = 0.67$ and 9.30\% at $z = 0.83$.

\subsection{Calibration of the D4000$_{\rm n} - z$ relation}
\label{calibration}

We use Eq. \ref{eqn:Hz}, rewritten in terms of D4000$_{\rm n}$: 
\begin{equation}
H(z) = - \frac{A(Z, M)}{1+z} \frac{dz}{dD4000_{\rm n}},
\end{equation}
 as also used in \citet{Moresco2012, Moresco2016, Loubser2025}, where $A(Z, M)$ (in units of Gyr$^{-1}$) is the conversion factor between age and D4000$_{\rm n}$, and $Z$ is stellar metallicity and $M$ a combination of stellar population modelling dependencies such as the assumed star formation history, stellar library, and stellar population models. The statistical uncertainty is entirely contained in $dz/dD4000_{\rm n}$ (the inverse of the slopes fitted in Figure \ref{fig:relation}), while the systematic uncertainty is entirely contained in $A(Z, M)$. 
 
\begin{figure}
\centering
\includegraphics[scale=0.38]{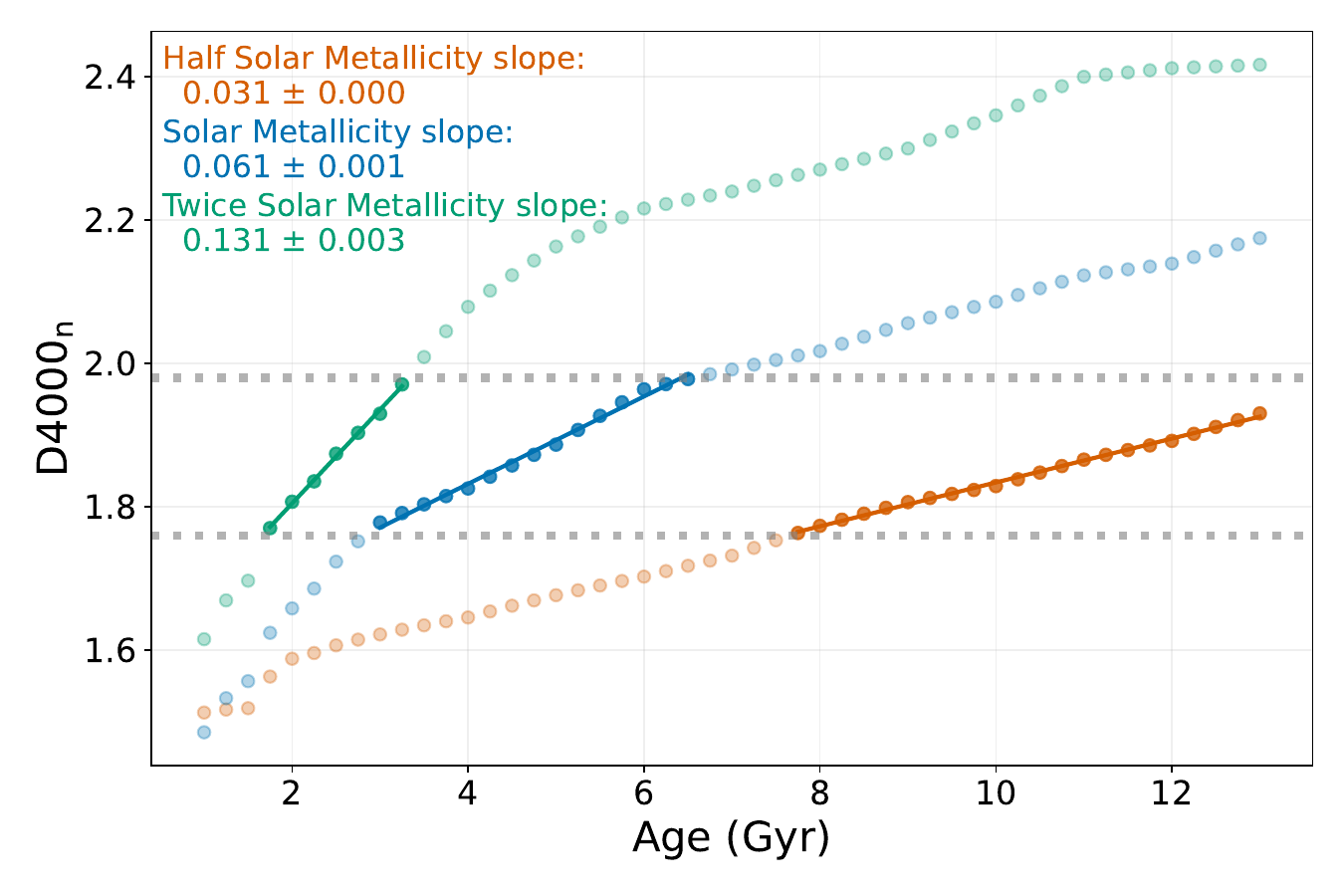} \\
\includegraphics[scale=0.46]{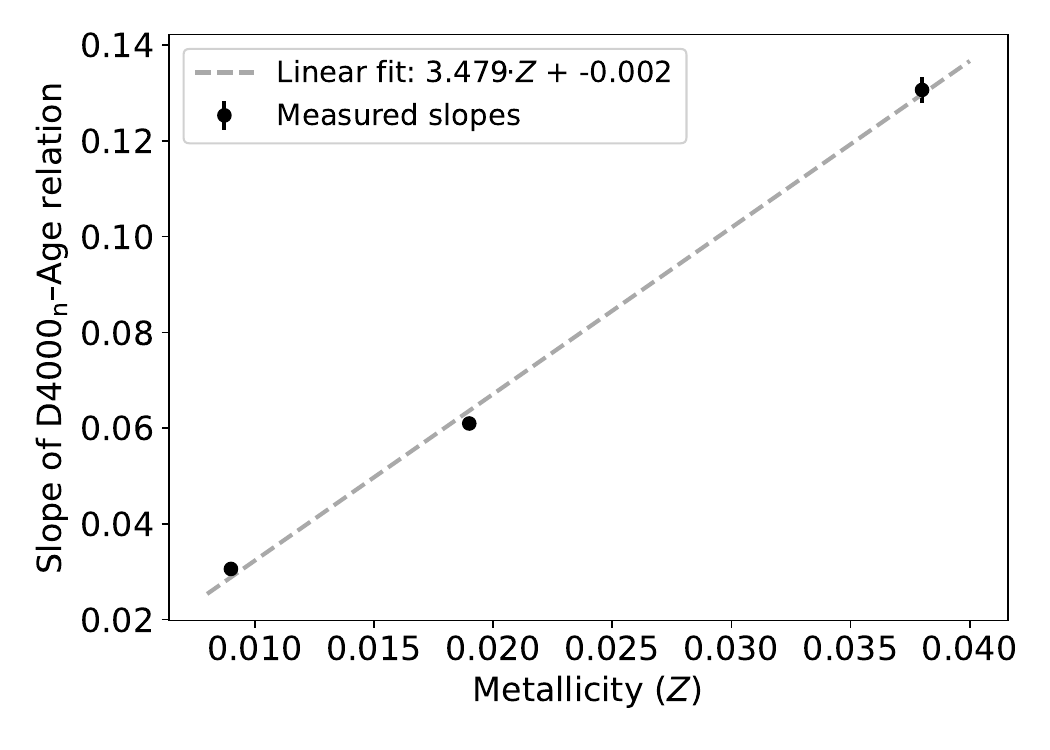} 
   \caption{The top panel shows MaStro \citep{Maraston2011} models with a Chabrier IMF and for its three stellar metallicities, $Z/Z_{\sun}$ of 0.5 ($Z = 0.009$, half solar), 1.0 ($Z = 0.019$, solar), and 2.0 ($Z = 0.038$, twice solar). The range for our median D4000$_{\rm n}$ values is indicated with the grey dashed horizontal lines and straight lines are fitted between these two lines. In the lower panel we show the three slopes, $A(Z, M)$, against the stellar metallicity.}
\label{fig:calibration}
\end{figure}

Our calibration of the D4000$_{\rm n} - z$ relation, to obtain $H(z)$, is a strong function of stellar metallicity ($Z$) through $A(Z, M)$. Instead of using the \texttt{FastSpecFit} fitted metallicity, which is heavily affected by the typical lack of features in the continuum spectra, and because fitted metallicities can be subject to the age--metallicity degeneracy, we rather derive the sensitivity of our $H(z)$ measurements to $Z$, as shown in Figure \ref{fig:calibration}. This also allows us to simultaneously understand the contribution of the stellar metallicity to the total systematic uncertainty. Here, we use the MaStro (M11) models \citep{Maraston2011}, with a resolution of 2.3 \AA{}, that have been constructed using the MILES \citep{Vazdekis2010} stellar library and a Chabrier IMF \citep{Chabrier2003}. These models are provided for three stellar metallicities, $Z/Z_{\sun}$ of 0.5, 1.0, and 2.0. We choose these particular models because we can fit one straight line in the region of our D4000$_{\rm n}$ measurements, for each of the three metallicities, in Figure \ref{fig:calibration} (top panel), but we incorporate the uncertainty due to the choice of model in our systematic errors in Section \ref{errors}. We indicate the range of median D4000$_{\rm n}$ measurements that we find for our sample (1.76 to 1.98) with two grey dashed lines in Figure \ref{fig:calibration} (top panel). 
 
From the literature, we expect the stellar metallicity of massive, passive galaxies to be well constrained around solar, or slightly over-solar, metallicity. In particular, for a large sample of nearby massive, passive galaxies from SDSS ($\log M_{\star}/M_{\sun} > 10.75$, and $z<0.3$), \citet{Citro2016} found $Z \sim 0.027 \pm 0.002$ (with a slight variation depending on the stellar synthesis models used), confirming earlier studies e.g. \citet{Gallazzi2005}. Some studies report no significant evolution in stellar metallicity for massive, passive galaxies with redshift (e.g. \citealt{Gallazzi2014, Borghi2022a, Jiao2023}, up to $z \sim 0.7$; and \citealt{Onodera2015, Estrada-Carpenter2019}, up to $z \sim 1.6$). On the other hand, other studies, e.g., \citet{Beverage2021}, report a non-negligible evolution ($\sim$ 0.1 to 0.2 dex) since $z=0.7$, with nearby galaxies being more metal rich than their higher redshift counterparts. However, we do not expect the mean stellar metallicity of our galaxies ($0.3 < z < 1.0$) to deviate too much from solar metallicity. In the lower panel of Figure \ref{fig:calibration}, we show the three fitted slopes, $A(Z, M)$ from the top panel, as a function of stellar metallicity. Here, too, we can fit a straight line. 

When we assume a typical uncertainty of 10\% on a fitted $Z$ (around solar metallicity), we can use this straight line and an analytical linear error propagation function and find that it propagates to an uncertainty of $\sim$10\% on $A(Z, M)$, and therefore on the systematic uncertainty on $H(z)$. This agrees with the most conservative uncertainty contribution (that is, the worst-case scenario, 9.8\%) derived in \citet{Moresco2020}, whereas the metallicity only contributes 1.9\% to the total uncertainty in their best-case scenario. We also test the fractional shift in $H(z)$ when other models are used to derive the sensitivity to $Z$, e.g., for the FSPS models \citep{Conroy2009, Conroy2010}, using the MILES library and Chabrier IMF, we find a fractional shift in $H(z)$ of 6.5\% (at solar metallicity), compared to when the MaStro models are used. This agrees with the uncertainty contribution of the stellar population models described and used in Section \ref{errors}.

\subsection{Systematic and total errors}
\label{errors}

Various previous CC studies \citep{Moresco2020, Moresco2022, Borghi2022b} have been dedicated to derive the contributions of different components, related to the stellar population modelling, to the systematic uncertainty on $H(z)$. The same equation as used in \citet{Loubser2025}, derived from the analysis by \citet{Moresco2020}, for a conservative systematic error ($\sigma_{\rm syst}$) is applicable here, the only difference being the contribution of the stellar metallicity. The analysis by \citet{Moresco2020} cross-check and validate five stellar population models, including the MaStro \citep{Maraston2011} models, which we use here for the derivation of the $H(z)$ to stellar metallicity sensitivity (Section \ref{calibration}). Our DESI DR1 CCs has D4000$_{\rm n}$ measurements well within the range tested in \citet{Moresco2020}. The contribution of the metallicity component of 3\% in \citet{Loubser2025} was specifically derived for the use of BCGs as CCs (which had higher D4000$_{\rm n}$ measurements), whereas the other contributions remain the same. We account for the typically slightly lower stellar metallicity of all massive, passive galaxies (compared to only BCGs), and use the slopes fitted in Figure \ref{fig:calibration}. For the BCGs, we measured D4000$_{\rm n}$ between 1.9 and 2.3. Here, because we measure the median D4000$_{\rm n}$ values from 1.76 to 1.98, where the slope of $A(Z, M)$ is steeper, it leads to a greater contribution to the systematic uncertainty. As shown in Section \ref{calibration}, the contribution of stellar metallicity to the systematic uncertainty in the case of this sample of massive passive galaxies is 10\%. Thus, we can adapt the equation for $\sigma_{\rm syst}$ to:
\begin{equation}
\sigma_{\rm syst} = \pm 3 \% (\rm SFH) \pm 7 \% (\rm library) \pm 6 \% (\rm model) \pm 10 \% (Z),
\end{equation} 
where the four contributions are derived from the choices of star formation history (SFH), stellar library, stellar population model, and stellar metallicity ($Z$). The contributions are added in quadrature. Thus, the systematic error amounts to 13.9\% (as opposed to 10.1\% in the case where only BCGs are used). 
 
In the redshift range $0.3 < z < 0.55$, where $dD4000_{\rm n}/dz$ is --0.426 $\pm$ 0.003, and with solar metallicity, we find:
\begin{equation}
H(z)= 88.48 \pm\ 0.57(\rm stat) \pm 12.32(\rm syst) \rm\ km\ s^{-1}\ Mpc^{-1} at\ \textit{z}=0.46.
\end{equation}
In the $0.55 < z < 0.78$ range, where $dD4000_{\rm n}/dz$ is --0.299 $\pm$ 0.016, and with solar metallicity, we find:
\begin{equation}
H(z)=119.45 \pm\ 6.39(\rm stat) \pm 16.64(\rm syst) \rm\ km\ s^{-1}\ Mpc^{-1} at\ \textit{z}=0.67.
\end{equation}
Lastly, in the $0.78 < z < 1.0$ range, where $dD4000_{\rm n}/dz$ is --0.301 $\pm$ 0.028, and with solar metallicity, we find:
\begin{equation}
H(z)=108.28 \pm\ 10.07(\rm stat) \pm 15.08(\rm syst) \rm\ km\ s^{-1}\ Mpc^{-1} at\ \textit{z}=0.83.
\end{equation}

The systematic uncertainties are common to the three $H(z)$ measurements, and highly correlated. For cosmological fitting, combining our data with other data, and accurate uncertainty propagation, we provide the correlation coefficients between the three measurements: $\rho$($z$=0.46, $z$=0.67) = 0.932, $\rho$($z$=0.46, $z$=0.83) = 0.830, and $\rho$($z$=0.67, $z$=0.83) = 0.776. 

\section{Our measurements in a cosmological context}
\subsection{The evolution of $H(z)$}
\label{comparisons}

\begin{figure}
\centering
\includegraphics[scale=0.58]{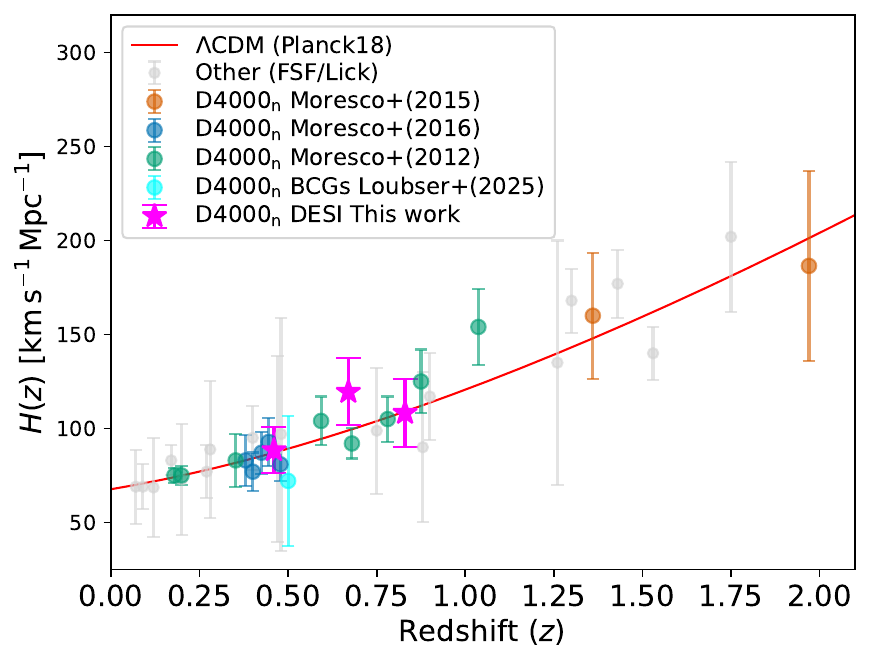} \\
\includegraphics[scale=0.58]{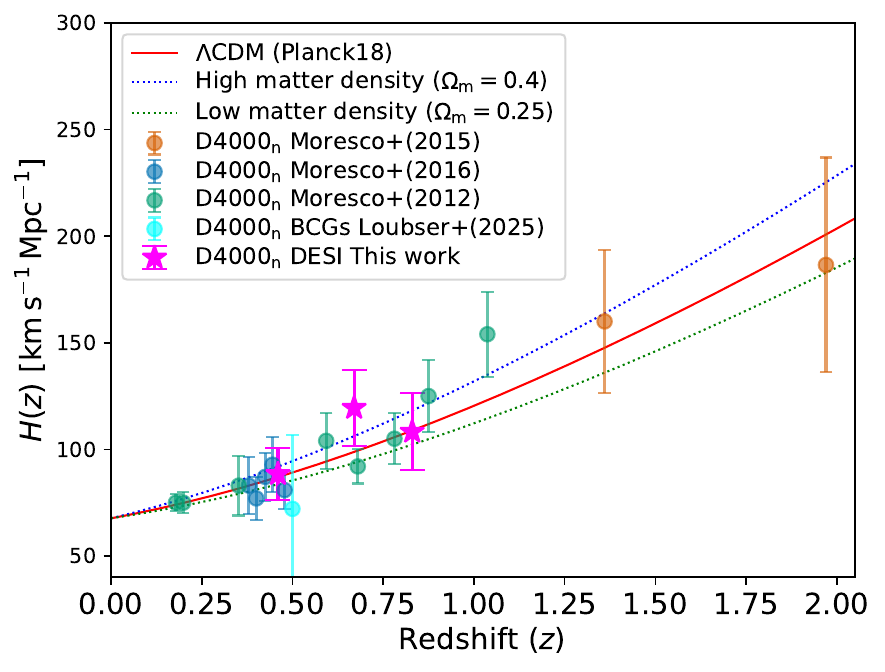}
   \caption{Top: $H(z)$ derived from CCs at different redshifts. We indicate measurements made using full-spectrum fitting (FSF) or Lick indices as light grey data points, and emphasize measurements made with D4000$_{\rm n}$ in colour. To guide the eye, we show the standard $\Lambda$CDM \texttt{Planck18} $H(z)$ evolution in a solid red line. We show our new measurements in magenta. Bottom: We only show the CC D4000$_{\rm n}$ measurements, and we show a simple extension of flat $\Lambda$CDM with the matter density set as $\Omega_{\rm m}=0.4$ (high matter density) shown with a blue dotted line, as well as a $\Omega_{\rm m}=0.25$ (low matter density) with a green dotted line, to illustrate possible applications.}
\label{fig:Hz}
\end{figure}

We compare our measurements with previous measurements made with CC in Figure \ref{fig:Hz} (top panel). We distinguish between CC measurements made using full-spectrum fitting (FSF) or Lick indices by indicating them in grey (a compilation of data from \citealt{Ratsimbazafy2017, Borghi2022b, Jimenez2023, Jiao2023, Tomasetti2023}), and we emphasise measurements made with D4000$_{\rm n}$ in colour (from \citealt{Moresco2012, Moresco2015, Moresco2016}, as well as \citealt{Loubser2025}, although it should be noted that the latter only uses BCGs). In the full-spectrum fitting approach, it is more difficult to accurately separate the various contributions to the systematic errors from the ingredients of the stellar population models and the degeneracies between them (see also the discussion in \citealt{Kjerrgren2023}). For convenience, we summarise all the CC measurements made using D4000$_{\rm n}$ in the Appendix \ref{OHDD4000}. It constitutes a homogeneous set of CC observational Hubble data points (OHDs) to test cosmological models. We add our three new measurements made from DESI DR1 in Figure \ref{fig:Hz} in magenta. As is standard in  CC $H(z)$ evolution plots, we also show a reference $\Lambda$CDM $H(z)$ evolution, for comparison (solid red line). This was obtained using the built-in \texttt{Planck18} cosmology in the \texttt{astropy.cosmology} module. It sets $H_{0}=67.66 \rm\ km\ s^{-1}\ Mpc^{-1}$, $\Omega_{\rm m}=0.30966$ and $\Omega_{\Lambda}=0.69034$, where $\Omega_{\rm m}$ is the fractional matter density and $\Omega_{\Lambda}$ is the fractional density of the cosmological constant \citep{Planck2018params, Planck2018_para, Astropy2013, Astropy2018}. 

While not the main purpose of this paper, we show two additional cosmological models in the bottom panel of Figure \ref{fig:Hz} for interest, and to illustrate the possible applications to cosmology. Here we show only the CC measurements made using D4000$_{\rm n}$ as given in the Appendix \ref{OHDD4000}. The standard reference is the \texttt{Planck18} cosmology as described above and shown with the solid red line. We then show a simple extension of flat $\Lambda$CDM with the matter density set as $\Omega_{\rm m}=0.4$ (high matter density) shown with a blue dotted line, as well as a $\Omega_{\rm m}=0.25$ (low matter density) with a green dotted line. This illustrates how the $H(z)$ curve bends. A higher $\Omega_{\rm m}$ translates to a higher $H(z)$ at $z>0$ because the Universe was more matter-dominated and decelerating faster. 

For our three new $H(z)$ measurements shown in Figure \ref{fig:Hz}, we do a pull analysis and find that the $z=0.67$ data point is $1.2\sigma$ above the \textit{Planck} $\Lambda$CDM (with $\Omega_{\rm m}=0.30966$) cosmology, whereas the other points are within $0.1\sigma$, and with a $p$-value of 0.72, there is no significant tension with \texttt{Planck18} ($p$ > 0.05). While not significant in this case, a higher $H(z)$ could suggest a higher expansion rate \citep{Riess2022}, or a dark energy equation of state with $w < -1$, or a dark energy equation of state that is evolving. Dedicated, rigorous, systematic studies are needed to test cosmological models against OHD measurements. 

\subsection{Deprojection and $H_{0}$ probes}
\label{deprojection}

\begin{figure}
\centering
\includegraphics[scale=0.62, clip]{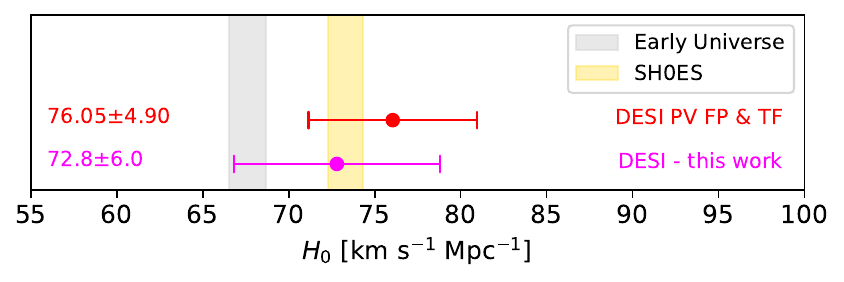}
   \caption{Our $H_{0}$ estimate (in magenta), for a flat $\Lambda$CDM with a fixed $\Omega_{\rm m}$, compared to the CMB early-universe and SH0ES measurements (as described in Section \ref{deprojection}). We also show the $H_{0}$ value derived by \citet{Said2025} using the DESI PV survey (using the Fundamental Plane and optical Tully–Fisher relation).}
\label{fig:H0}
\end{figure}	 

We now deproject our measurements of $H(z=0.46)$, $H(z=0.67)$ and $H(z=0.83)$ to $z=0$, and obtain an estimate of $H_0$ using
\begin{equation}
H(z) = H_{0} \sqrt{\Omega_{\rm m} (1 + z)^{3} + (1-\Omega_{\rm m})}, \label{eq:Hz-eq}
\end{equation}
for a flat $\Lambda$CDM with a fixed $\Omega_{\rm m}$. We use the \texttt{FlatLambdaCDM} module in \texttt{astropy.cosmology} with $\Omega_{\rm m}=0.30966$. For this illustration we add the statistical and systematic uncertainties in quadrature, and it amounts to 12.3 (13.9\%) at $z=0.46$, 17.8 (14.9\%) at $z=0.67$, and to 18.1 (16.7\%) at $z=0.83$. The results of the deprojection, respectively, is $H_{0}=68.74\pm 10.81$, $H_{0}=81.81\pm 10.20$, and ${H}_{0}=67.28\pm 10.38\ \rm km\ s^{-1}\ Mpc^{-1}$, where errors include systematic and statistical uncertainties. The deprojection to $H_{0}$ depends on the cosmological model that is used, however $\Omega_{\rm m}$ is extremely well constrained from various measurements. When deprojecting our $H(z)$ measurements to $H_{0}$ under the flat $\Lambda$CDM assumption with $\Omega_{\rm m}$ = 0.30966, we estimate an additional systematic uncertainty of $\pm$0.5--1.5 $\rm km\ s^{-1}\ Mpc^{-1}$ from varying $\Omega_{\rm m}$ within the conservative, \textit{Planck}-constrained range of 0.30 -- 0.32. This cosmological systematic is negligible compared to our dominant systematic uncertainties of $\pm$12--17 $\rm km\ s^{-1}\ Mpc^{-1}$. We can combine the three estimates into one $H_{0}$ using an inverse-variance weighted mean to find: 
 
\begin{equation}
H_{0}=72.82\pm 6.03\ \rm km\ s^{-1}\ Mpc^{-1}
\end{equation}
 
Figure \ref{fig:H0} shows our deprojected value of $H_{0}$ (in magenta) compared to a range determined by combining various recent early Universe measurements (specifically \citealt{Planck2018_para, Alam2021, Aiola2020, Madhavacheril2024, Abbott2022-DES, Adame2024EDR, Louis2025}), and indicated with a light grey vertical band. We also show the ``SH0ES'' late Universe measurement (in a light yellow vertical band), made using calibration from Hubble Space Telescope observations of Cepheid variables in supernova (SN) type Ia host galaxies \citep{Riess2022}. For comparison, we also include the $H_{0}$ value recently derived by \citet{Said2025} using the DESI peculiar velocity (PV) survey of early- and late-type galaxies within the DESI footprint using both the Fundamental Plane (FP) and optical Tully–Fisher (TF) relation ``DESI PV FP \& TF" in Figure \ref{fig:H0}. Our deprojected $H_{0}$ estimate agrees more closely with the late-Universe methods (as already suggested by Figure \ref{fig:Hz}), but within the total (systematic and statistical) uncertainties, it is consistent with both the late- and early-Universe measurements.

\section{Summary}
\label{summary}

The DESI survey is already delivering unprecedented volumes of galaxy spectroscopic data. DESI DR1 is nearly four times larger than all previous Sloan Digital Sky Survey (SDSS) programmes combined \citep{DESIDR1}. This provides the ability to measure constraints on cosmological tensions and to study galaxy evolution with huge statistical samples. \citet{Louis2025} presents an $H_{0}$ measurement using baryon acoustic oscillations (BAO) data from DESI, combined with the Atacama Cosmology Telescope (ACT) DR6 power spectra, and CMB lensing from ACT and \textit{Planck}, that agrees with other early-Universe measurements, consistent with $\Lambda$CDM. But, as the first three years of DESI data are analysed, there are mounting indications that the impact of dark energy may be weakening over time and that other models may be a better fit. It is therefore particularly interesting to analyse the DESI data using the Cosmic Chronometer (CC) method, which gives an independent measurement of $H(z)$, not assuming a cosmological model. CCs do not depend on early-time physics or the traditional cosmic distance ladder and are directly related to the value of $H(z)$ at a specific redshift, rather than to the integral of $H(z)$, which is commonly probed in angular diameter measurements.

We extracted all massive galaxies ($\log M_{\star}/M_{\sun} > 10.75$, and $\sigma$ > 280 km s$^{-1}$) without emission in [OII] $\lambda$3727 \AA{}, and with reliable redshifts and D4000$_{\rm n}$ measurements from DESI DR1. From this sample of 360 000 massive, passive galaxies, the largest CC sample yet, we use D4000$_{\rm n}$ to get a direct measurement of $H(z)$ with reliable statistical and systematic uncertainties. We find $H(z)=$ 88.48 $\pm\ 0.57(\rm stat) \pm 12.32(\rm syst)$, $H(z)=$ 119.45 $\pm\ 6.39(\rm stat) \pm 16.64(\rm syst)$, and $H(z)= 108.28 \pm 10.07(\rm stat) \pm 15.08(\rm syst)$ $\rm km\ s^{-1}\ Mpc^{-1}$ at $z=0.46$,  $z=0.67$, and $z=0.83$, respectively.

For our three new $H(z)$ measurements shown in Figure \ref{fig:Hz}, two are remarkably consistent with the \texttt{Planck18} $\Lambda$CDM reference evolution shown, while the $z=0.67$ data point is $1.2\sigma$ above the \texttt{Planck18} reference evolution. With a $p$-value of 0.72, there is no significant tension with \texttt{Planck18}. For convenience, we summarise a separate, homogeneous OHD data set using only D4000$_{\rm n}$ CC measurements that can be used to test cosmological models more rigorously in dedicated future studies. We can deproject our measurements to $H_{0}$ to compare it to other probes, and we find $H_{0}=72.82\pm 6.03\ \rm km\ s^{-1}\ Mpc^{-1}$. Our estimate agrees more closely with the late-Universe methods, but within the total (systematic and statistical) uncertainties, it is consistent with both the late- and early-Universe measurements. 

Our statistical uncertainties reach 0.65\%, 5.35\%, and 9.30\% on our three measurements. Now that this statistical precision is achievable, the main area for future improvements is the systematic uncertainties. To improve on the systematic uncertainties, while still using very large galaxy samples, we would need to improve the method's sensitivity to stellar metallicity by refining and improving the metallicity grids and ranges of the stellar population models. This will, in turn, also allow more accurate stellar metallicity measurements from galaxy spectra. Stellar population models are continuously being improved, e.g., to allow for $\alpha$-enhancements \citep{Vazdekis2015}, improved theoretical stellar libraries e.g., \texttt{C3K} in FSPS \citep{Conroy2010}, bayesian analysis and the inclusion of non-parametric star formation histories \citep{Johnson2021}. Current, and future surveys e.g., 4MOST \citep{deJong2022} promises larger and more precise empirical stellar libraries, and hopefully our understanding of the inconsistencies between different models will also improve. The alternative method is to use a more homogeneous sample selection of the most massive galaxies (BCGs), e.g., in \citet{Loubser2025}, but then the statistical uncertainty increases tremendously due to limited sample size. 

In our analysis, we also illustrate that even amongst samples of the massive, passive galaxies, the effect of downsizing can clearly be seen. We confirm results from previous CC studies (e.g., \citealt{Borghi2022b}), that only the highest velocity dispersion galaxies (> 280 km s$^{-1}$) should be used for accurate CC studies. As a consequence, the CC approach absolutely relies on accurate velocity dispersion or stellar mass measurements to identify massive, passive galaxies at any given redshift. 

\section*{Acknowledgements} 

We thank the reviewer for their suggestions that has improved the robustness of the results. This work is based on research supported in part by the National Research Foundation (NRF) of South Africa (NRF Grant Number: CPRR240414214079). Any opinion, finding, and conclusion or recommendation expressed in this material is that of the author(s), and the NRF does not accept any liability in this regard. 

This research used data obtained with the Dark Energy Spectroscopic Instrument (DESI). DESI construction and operations is managed by the Lawrence Berkeley National Laboratory. This material is based upon work supported by the U.S. Department of Energy, Office of Science, Office of High-Energy Physics, under Contract No. DE–AC02–05CH11231, and by the National Energy Research Scientific Computing Center, a DOE Office of Science User Facility under the same contract. Additional support for DESI was provided by the U.S. National Science Foundation (NSF), Division of Astronomical Sciences under Contract No. AST-0950945 to the NSF’s National Optical-Infrared Astronomy Research Laboratory; the Science and Technology Facilities Council of the United Kingdom; the Gordon and Betty Moore Foundation; the Heising-Simons Foundation; the French Alternative Energies and Atomic Energy Commission (CEA); the National Council of Humanities, Science and Technology of Mexico (CONAHCYT); the Ministry of Science and Innovation of Spain (MICINN), and by the DESI Member Institutions: www.desi.lbl.gov/collaborating-institutions. The DESI collaboration is honored to be permitted to conduct scientific research on I’oligam Du’ag (Kitt Peak), a mountain with particular significance to the Tohono O’odham Nation. Any opinions, findings, and conclusions or recommendations expressed in this material are those of the author(s) and do not necessarily reflect the views of the U.S. National Science Foundation, the U.S. Department of Energy, or any of the listed funding agencies.

This research used Astropy,\footnote{\url{http://www.astropy.org}} a community-developed core Python package for Astronomy \citep{Astropy2013, Astropy2018}, as well as $\mathtt{pwlf}$\ \citep{pwlf}. We also make use of the \texttt{FastSpecFit} VAC created by \citet{fastspecfit}, and the \texttt{Marvin} webservice and python tools (\url{https://magrathea.sdss.org/marvin/}). 

\section*{Data availability}

All data underlying this article are publicly available. DESI DR1 data are publicly available as described in \citet{DESIDR1}, and the VAC used is publicly available at \url{https://data.desi.lbl.gov/public/dr1/vac/dr1/fastspecfit/iron/v3.0/catalogs/}. We also use the publicly-available MaNGA \texttt{Firefly} VAC (\url{https://www.sdss4.org/dr16/manga/manga-data/manga-firefly-value-added-catalog/}), and MaNGA DR16 flux cubes in the Appendix \ref{apertures}. 



\bibliographystyle{mnras}
\bibliography{References} 

\begin{thebibliography}{}
\makeatletter
\relax
\def\mn@urlcharsother{\let\do\@makeother \do\$\do\&\do\#\do\^\do\_\do\%\do\~}
\def\mn@doi{\begingroup\mn@urlcharsother \@ifnextchar [ {\mn@doi@}
  {\mn@doi@[]}}
\def\mn@doi@[#1]#2{\def\@tempa{#1}\ifx\@tempa\@empty \href
  {http://dx.doi.org/#2} {doi:#2}\else \href {http://dx.doi.org/#2} {#1}\fi
  \endgroup}
\def\mn@eprint#1#2{\mn@eprint@#1:#2::\@nil}
\def\mn@eprint@arXiv#1{\href {http://arxiv.org/abs/#1} {{\tt arXiv:#1}}}
\def\mn@eprint@dblp#1{\href {http://dblp.uni-trier.de/rec/bibtex/#1.xml}
  {dblp:#1}}
\def\mn@eprint@#1:#2:#3:#4\@nil{\def\@tempa {#1}\def\@tempb {#2}\def\@tempc
  {#3}\ifx \@tempc \@empty \let \@tempc \@tempb \let \@tempb \@tempa \fi \ifx
  \@tempb \@empty \def\@tempb {arXiv}\fi \@ifundefined
  {mn@eprint@\@tempb}{\@tempb:\@tempc}{\expandafter \expandafter \csname
  mn@eprint@\@tempb\endcsname \expandafter{\@tempc}}}

\bibitem[\protect\citeauthoryear{{Abbott} et~al.,}{{Abbott}
  et~al.}{2022}]{Abbott2022-DES}
{Abbott} T.~M.~C.,  et~al., 2022, \mn@doi [Physics Review D]
  {10.1103/PhysRevD.105.023520}, \href
  {https://ui.adsabs.harvard.edu/abs/2022PhRvD.105b3520A} {105, 023520}

\bibitem[\protect\citeauthoryear{{Ahlstr{\"o}m Kjerrgren} \&
  {M{\"o}rtsell}}{{Ahlstr{\"o}m Kjerrgren} \&
  {M{\"o}rtsell}}{2023}]{Kjerrgren2023}
{Ahlstr{\"o}m Kjerrgren} A.,  {M{\"o}rtsell} E.,  2023, \mn@doi [\mnras]
  {10.1093/mnras/stac1978}, \href
  {https://ui.adsabs.harvard.edu/abs/2023MNRAS.518..585A} {518, 585}

\bibitem[\protect\citeauthoryear{{Aiola} et~al.,}{{Aiola}
  et~al.}{2020}]{Aiola2020}
{Aiola} S.,  et~al., 2020, \mn@doi [JCAP] {10.1088/1475-7516/2020/12/047},
  \href {https://ui.adsabs.harvard.edu/abs/2020JCAP...12..047A} {2020, 047}

\bibitem[\protect\citeauthoryear{Alam, Aubert, Avila  \& et al.}{Alam
  et~al.}{2021}]{Alam2021}
Alam S.,  Aubert M.,  Avila S.,   et al. 2021, \mn@doi [Phys. Rev. D]
  {10.1103/PhysRevD.103.083533}, 103, 083533

\bibitem[\protect\citeauthoryear{{Astropy Collaboration} et~al.,}{{Astropy
  Collaboration} et~al.}{2013}]{Astropy2013}
{Astropy Collaboration} et~al., 2013, \mn@doi [A\&A]
  {10.1051/0004-6361/201322068}, \href
  {https://ui.adsabs.harvard.edu/abs/2013A&A...558A..33A} {558, A33}

\bibitem[\protect\citeauthoryear{{Astropy Collaboration} et~al.,}{{Astropy
  Collaboration} et~al.}{2018}]{Astropy2018}
{Astropy Collaboration} et~al., 2018, \mn@doi [AJ] {10.3847/1538-3881/aabc4f},
  \href {https://ui.adsabs.harvard.edu/abs/2018AJ....156..123A} {156, 123}

\bibitem[\protect\citeauthoryear{{Balogh}, {Morris}, {Yee}, {Carlberg}  \&
  {Ellingson}}{{Balogh} et~al.}{1999}]{Balogh1999}
{Balogh} M.~L.,  {Morris} S.~L.,  {Yee} H. K.~C.,  {Carlberg} R.~G.,
  {Ellingson} E.,  1999, \mn@doi [ApJ] {10.1086/308056}, \href
  {http://saaoads.chpc.ac.za/abs/1999ApJ...527...54B} {527, 54}

\bibitem[\protect\citeauthoryear{{Beverage}, {Kriek}, {Conroy}, {Bezanson},
  {Franx}  \& {van der Wel}}{{Beverage} et~al.}{2021}]{Beverage2021}
{Beverage} A.~G.,  {Kriek} M.,  {Conroy} C.,  {Bezanson} R.,  {Franx} M.,
  {van der Wel} A.,  2021, \mn@doi [\apjl] {10.3847/2041-8213/ac12cd}, \href
  {https://ui.adsabs.harvard.edu/abs/2021ApJ...917L...1B} {917, L1}

\bibitem[\protect\citeauthoryear{{Borghi}, {Moresco}, {Cimatti}, {Huchet},
  {Quai}  \& {Pozzetti}}{{Borghi} et~al.}{2022a}]{Borghi2022a}
{Borghi} N.,  {Moresco} M.,  {Cimatti} A.,  {Huchet} A.,  {Quai} S.,
  {Pozzetti} L.,  2022a, \mn@doi [ApJ] {10.3847/1538-4357/ac3240}, \href
  {https://ui.adsabs.harvard.edu/abs/2022ApJ...927..164B} {927, 164}

\bibitem[\protect\citeauthoryear{{Borghi}, {Moresco}  \& {Cimatti}}{{Borghi}
  et~al.}{2022b}]{Borghi2022b}
{Borghi} N.,  {Moresco} M.,   {Cimatti} A.,  2022b, \mn@doi [ApJL]
  {10.3847/2041-8213/ac3fb2}, \href
  {https://ui.adsabs.harvard.edu/abs/2022ApJ...928L...4B} {928, L4}

\bibitem[\protect\citeauthoryear{{Bundy} et~al.,}{{Bundy}
  et~al.}{2015}]{Bundy2015}
{Bundy} K.,  et~al., 2015, \mn@doi [\apj] {10.1088/0004-637X/798/1/7}, \href
  {https://ui.adsabs.harvard.edu/abs/2015ApJ...798....7B} {798, 7}

\bibitem[\protect\citeauthoryear{{Chabrier}}{{Chabrier}}{2003}]{Chabrier2003}
{Chabrier} G.,  2003, \mn@doi [\pasp] {10.1086/376392}, \href
  {https://ui.adsabs.harvard.edu/abs/2003PASP..115..763C} {115, 763}

\bibitem[\protect\citeauthoryear{{Citro}, {Pozzetti}, {Moresco}  \&
  {Cimatti}}{{Citro} et~al.}{2016}]{Citro2016}
{Citro} A.,  {Pozzetti} L.,  {Moresco} M.,   {Cimatti} A.,  2016, \mn@doi
  [\aap] {10.1051/0004-6361/201527772}, \href
  {https://ui.adsabs.harvard.edu/abs/2016A&A...592A..19C} {592, A19}

\bibitem[\protect\citeauthoryear{{Clerici}, {Schnorr-M{\"u}ller}, {Trevisan}
  \& {Ricci}}{{Clerici} et~al.}{2024}]{Clerici2024}
{Clerici} K.~S.,  {Schnorr-M{\"u}ller} A.,  {Trevisan} M.,   {Ricci} T.~V.,
  2024, \mn@doi [\mnras] {10.1093/mnras/stae1213}, \href
  {https://ui.adsabs.harvard.edu/abs/2024MNRAS.531.1034C} {531, 1034}

\bibitem[\protect\citeauthoryear{{Conroy} \& {Gunn}}{{Conroy} \&
  {Gunn}}{2010}]{Conroy2010}
{Conroy} C.,  {Gunn} J.~E.,  2010, \mn@doi [\apj]
  {10.1088/0004-637X/712/2/833}, \href
  {https://ui.adsabs.harvard.edu/abs/2010ApJ...712..833C} {712, 833}

\bibitem[\protect\citeauthoryear{{Conroy}, {Gunn}  \& {White}}{{Conroy}
  et~al.}{2009}]{Conroy2009}
{Conroy} C.,  {Gunn} J.~E.,   {White} M.,  2009, \mn@doi [\apj]
  {10.1088/0004-637X/699/1/486}, \href
  {https://ui.adsabs.harvard.edu/abs/2009ApJ...699..486C} {699, 486}

\bibitem[\protect\citeauthoryear{{Cowie}, {Songaila}, {Hu}  \& {Cohen}}{{Cowie}
  et~al.}{1996}]{Cowie1996}
{Cowie} L.~L.,  {Songaila} A.,  {Hu} E.~M.,   {Cohen} J.~G.,  1996, \mn@doi
  [\aj] {10.1086/118058}, \href
  {https://ui.adsabs.harvard.edu/abs/1996AJ....112..839C} {112, 839}

\bibitem[\protect\citeauthoryear{{DESI Collaboration} et~al.,}{{DESI
  Collaboration} et~al.}{2024}]{Adame2024EDR}
{DESI Collaboration} et~al., 2024, \mn@doi [\aj] {10.3847/1538-3881/ad3217},
  \href {https://ui.adsabs.harvard.edu/abs/2024AJ....168...58D} {168, 58}

\bibitem[\protect\citeauthoryear{{DESI Collaboration} et~al.,}{{DESI
  Collaboration} et~al.}{2025}]{DESIDR1}
{DESI Collaboration} et~al., 2025, \mn@doi [arXiv e-prints]
  {10.48550/arXiv.2503.14745}, \href
  {https://ui.adsabs.harvard.edu/abs/2025arXiv250314745D} {p. arXiv:2503.14745}

\bibitem[\protect\citeauthoryear{{Dey} et~al.,}{{Dey} et~al.}{2019}]{Dey2019}
{Dey} A.,  et~al., 2019, \mn@doi [\aj] {10.3847/1538-3881/ab089d}, \href
  {https://ui.adsabs.harvard.edu/abs/2019AJ....157..168D} {157, 168}

\bibitem[\protect\citeauthoryear{{Di Valentino} et~al.,}{{Di Valentino}
  et~al.}{2021}]{DiValentino2021}
{Di Valentino} E.,  et~al., 2021, \mn@doi [Classical and Quantum Gravity]
  {10.1088/1361-6382/ac086d}, \href
  {https://ui.adsabs.harvard.edu/abs/2021CQGra..38o3001D} {38, 153001}

\bibitem[\protect\citeauthoryear{{Estrada-Carpenter}
  et~al.,}{{Estrada-Carpenter} et~al.}{2019}]{Estrada-Carpenter2019}
{Estrada-Carpenter} V.,  et~al., 2019, \mn@doi [\apj]
  {10.3847/1538-4357/aaf22e}, \href
  {https://ui.adsabs.harvard.edu/abs/2019ApJ...870..133E} {870, 133}

\bibitem[\protect\citeauthoryear{{Furlong} et~al.,}{{Furlong}
  et~al.}{2015}]{Furlong2015}
{Furlong} M.,  et~al., 2015, \mn@doi [\mnras] {10.1093/mnras/stv852}, \href
  {https://ui.adsabs.harvard.edu/abs/2015MNRAS.450.4486F} {450, 4486}

\bibitem[\protect\citeauthoryear{{Gallazzi}, {Charlot}, {Brinchmann}, {White}
  \& {Tremonti}}{{Gallazzi} et~al.}{2005}]{Gallazzi2005}
{Gallazzi} A.,  {Charlot} S.,  {Brinchmann} J.,  {White} S. D.~M.,   {Tremonti}
  C.~A.,  2005, \mn@doi [\mnras] {10.1111/j.1365-2966.2005.09321.x}, \href
  {https://ui.adsabs.harvard.edu/abs/2005MNRAS.362...41G} {362, 41}

\bibitem[\protect\citeauthoryear{{Gallazzi}, {Bell}, {Zibetti}, {Brinchmann}
  \& {Kelson}}{{Gallazzi} et~al.}{2014}]{Gallazzi2014}
{Gallazzi} A.,  {Bell} E.~F.,  {Zibetti} S.,  {Brinchmann} J.,   {Kelson}
  D.~D.,  2014, \mn@doi [\apj] {10.1088/0004-637X/788/1/72}, \href
  {https://ui.adsabs.harvard.edu/abs/2014ApJ...788...72G} {788, 72}

\bibitem[\protect\citeauthoryear{{Goddard} et~al.,}{{Goddard}
  et~al.}{2017}]{Goddard2017}
{Goddard} D.,  et~al., 2017, \mn@doi [\mnras] {10.1093/mnras/stw3371}, \href
  {https://ui.adsabs.harvard.edu/abs/2017MNRAS.466.4731G} {466, 4731}

\bibitem[\protect\citeauthoryear{{Guy} et~al.,}{{Guy} et~al.}{2023}]{Guy2023}
{Guy} J.,  et~al., 2023, \mn@doi [\aj] {10.3847/1538-3881/acb212}, \href
  {https://ui.adsabs.harvard.edu/abs/2023AJ....165..144G} {165, 144}

\bibitem[\protect\citeauthoryear{Jekel \& Venter}{Jekel \& Venter}{2019}]{pwlf}
Jekel C.~F.,  Venter G.,  2019, {pwlf:} A Python Library for Fitting 1D
  Continuous Piecewise Linear Functions.
\url {https://github.com/cjekel/piecewise_linear_fit_py}

\bibitem[\protect\citeauthoryear{{Jiao}, {Borghi}, {Moresco}  \&
  {Zhang}}{{Jiao} et~al.}{2023}]{Jiao2023}
{Jiao} K.,  {Borghi} N.,  {Moresco} M.,   {Zhang} T.-J.,  2023, \mn@doi [ApJS]
  {10.3847/1538-4365/acbc77}, \href
  {https://ui.adsabs.harvard.edu/abs/2023ApJS..265...48J} {265, 48}

\bibitem[\protect\citeauthoryear{{Jimenez} \& {Loeb}}{{Jimenez} \&
  {Loeb}}{2002}]{Jimenez2002}
{Jimenez} R.,  {Loeb} A.,  2002, \mn@doi [ApJ] {10.1086/340549}, \href
  {https://ui.adsabs.harvard.edu/abs/2002ApJ...573...37J} {573, 37}

\bibitem[\protect\citeauthoryear{{Jimenez}, {Moresco}, {Verde}  \&
  {Wandelt}}{{Jimenez} et~al.}{2023}]{Jimenez2023}
{Jimenez} R.,  {Moresco} M.,  {Verde} L.,   {Wandelt} B.~D.,  2023, \mn@doi
  [JCAP] {10.1088/1475-7516/2023/11/047}, \href
  {https://ui.adsabs.harvard.edu/abs/2023JCAP...11..047J} {2023, 047}

\bibitem[\protect\citeauthoryear{{Johnson}, {Leja}, {Conroy}  \&
  {Speagle}}{{Johnson} et~al.}{2021}]{Johnson2021}
{Johnson} B.~D.,  {Leja} J.,  {Conroy} C.,   {Speagle} J.~S.,  2021, \mn@doi
  [\apjs] {10.3847/1538-4365/abef67}, \href
  {https://ui.adsabs.harvard.edu/abs/2021ApJS..254...22J} {254, 22}

\bibitem[\protect\citeauthoryear{{Loubser}, {Babul}, {Hoekstra}, {Mahdavi},
  {Donahue}, {Bildfell}  \& {Voit}}{{Loubser} et~al.}{2016}]{Loubser2016}
{Loubser} S.~I.,  {Babul} A.,  {Hoekstra} H.,  {Mahdavi} A.,  {Donahue} M.,
  {Bildfell} C.,   {Voit} G.~M.,  2016, \mn@doi [MNRAS]
  {10.1093/mnras/stv2784}, \href
  {http://adsabs.harvard.edu/abs/2016MNRAS.456.1565L} {456, 1565}

\bibitem[\protect\citeauthoryear{{Loubser} et~al.,}{{Loubser}
  et~al.}{2024}]{Loubser2024}
{Loubser} S.~I.,  et~al., 2024, \mn@doi [MNRAS] {10.1093/mnras/stad3654}, \href
  {https://ui.adsabs.harvard.edu/abs/2024MNRAS.527.7158L} {527, 7158}

\bibitem[\protect\citeauthoryear{{Loubser} et~al.,}{{Loubser}
  et~al.}{2025}]{Loubser2025}
{Loubser} S.~I.,  et~al., 2025, \mn@doi [\mnras] {10.1093/mnras/staf915}, \href
  {https://ui.adsabs.harvard.edu/abs/2025MNRAS.540.3135L} {540, 3135}

\bibitem[\protect\citeauthoryear{{Louis} et~al.,}{{Louis}
  et~al.}{2025}]{Louis2025}
{Louis} T.,  et~al., 2025, \mn@doi [arXiv e-prints]
  {10.48550/arXiv.2503.14452}, \href
  {https://ui.adsabs.harvard.edu/abs/2025arXiv250314452L} {p. arXiv:2503.14452}

\bibitem[\protect\citeauthoryear{{Madhavacheril} et~al.,}{{Madhavacheril}
  et~al.}{2024}]{Madhavacheril2024}
{Madhavacheril} M.~S.,  et~al., 2024, \mn@doi [ApJ] {10.3847/1538-4357/acff5f},
  \href {https://ui.adsabs.harvard.edu/abs/2024ApJ...962..113M} {962, 113}

\bibitem[\protect\citeauthoryear{{Maraston} \& {Str{\"o}mb{\"a}ck}}{{Maraston}
  \& {Str{\"o}mb{\"a}ck}}{2011}]{Maraston2011}
{Maraston} C.,  {Str{\"o}mb{\"a}ck} G.,  2011, \mn@doi [MNRAS]
  {10.1111/j.1365-2966.2011.19738.x}, \href
  {http://saaoads.chpc.ac.za/abs/2011MNRAS.418.2785M} {418, 2785}

\bibitem[\protect\citeauthoryear{{Moresco}}{{Moresco}}{2015}]{Moresco2015}
{Moresco} M.,  2015, \mn@doi [\mnras] {10.1093/mnrasl/slv037}, \href
  {https://ui.adsabs.harvard.edu/abs/2015MNRAS.450L..16M} {450, L16}

\bibitem[\protect\citeauthoryear{{Moresco}}{{Moresco}}{2024}]{Moresco2024}
{Moresco} M.,  2024, \mn@doi [arXiv e-prints] {10.48550/arXiv.2412.01994},
  \href {https://ui.adsabs.harvard.edu/abs/2024arXiv241201994M} {p.
  arXiv:2412.01994}

\bibitem[\protect\citeauthoryear{{Moresco} et~al.,}{{Moresco}
  et~al.}{2012}]{Moresco2012}
{Moresco} M.,  et~al., 2012, \mn@doi [JCAP] {10.1088/1475-7516/2012/08/006},
  \href {https://ui.adsabs.harvard.edu/abs/2012JCAP...08..006M} {2012, 006}

\bibitem[\protect\citeauthoryear{{Moresco} et~al.,}{{Moresco}
  et~al.}{2016}]{Moresco2016}
{Moresco} M.,  et~al., 2016, \mn@doi [JCAP] {10.1088/1475-7516/2016/05/014},
  \href {https://ui.adsabs.harvard.edu/abs/2016JCAP...05..014M} {2016, 014}

\bibitem[\protect\citeauthoryear{{Moresco}, {Jimenez}, {Verde}, {Pozzetti},
  {Cimatti}  \& {Citro}}{{Moresco} et~al.}{2018}]{Moresco2018}
{Moresco} M.,  {Jimenez} R.,  {Verde} L.,  {Pozzetti} L.,  {Cimatti} A.,
  {Citro} A.,  2018, \mn@doi [ApJ] {10.3847/1538-4357/aae829}, \href
  {https://ui.adsabs.harvard.edu/abs/2018ApJ...868...84M} {868, 84}

\bibitem[\protect\citeauthoryear{{Moresco}, {Jimenez}, {Verde}, {Cimatti}  \&
  {Pozzetti}}{{Moresco} et~al.}{2020}]{Moresco2020}
{Moresco} M.,  {Jimenez} R.,  {Verde} L.,  {Cimatti} A.,   {Pozzetti} L.,
  2020, \mn@doi [ApJ] {10.3847/1538-4357/ab9eb0}, \href
  {https://ui.adsabs.harvard.edu/abs/2020ApJ...898...82M} {898, 82}

\bibitem[\protect\citeauthoryear{{Moresco} et~al.,}{{Moresco}
  et~al.}{2022}]{Moresco2022}
{Moresco} M.,  et~al., 2022, \mn@doi [Living Reviews in Relativity]
  {10.1007/s41114-022-00040-z}, \href
  {https://ui.adsabs.harvard.edu/abs/2022LRR....25....6M} {25, 6}

\bibitem[\protect\citeauthoryear{{Moustakas}, {Scholte}, {Dey}  \&
  {Khederlarian}}{{Moustakas} et~al.}{2023a}]{fastspecfit}
{Moustakas} J.,  {Scholte} D.,  {Dey} B.,   {Khederlarian} A.,  2023a,
  {FastSpecFit: Fast spectral synthesis and emission-line fitting of DESI
  spectra}, Astrophysics Source Code Library, record ascl:2308.005 (\mn@eprint
  {ascl} {2308.005})

\bibitem[\protect\citeauthoryear{{Moustakas} et~al.,}{{Moustakas}
  et~al.}{2023b}]{Moustakas2023}
{Moustakas} J.,  et~al., 2023b, \mn@doi [\apjs] {10.3847/1538-4365/acfaa2},
  \href {https://ui.adsabs.harvard.edu/abs/2023ApJS..269....3M} {269, 3}

\bibitem[\protect\citeauthoryear{{Onodera} et~al.,}{{Onodera}
  et~al.}{2015}]{Onodera2015}
{Onodera} M.,  et~al., 2015, \mn@doi [\apj] {10.1088/0004-637X/808/2/161},
  \href {https://ui.adsabs.harvard.edu/abs/2015ApJ...808..161O} {808, 161}

\bibitem[\protect\citeauthoryear{{Parikh} et~al.,}{{Parikh}
  et~al.}{2018}]{Parikh2018}
{Parikh} T.,  et~al., 2018, \mn@doi [\mnras] {10.1093/mnras/sty785}, \href
  {https://ui.adsabs.harvard.edu/abs/2018MNRAS.477.3954P} {477, 3954}

\bibitem[\protect\citeauthoryear{{Planck Collaboration}}{{Planck
  Collaboration}}{2020}]{Planck2018params}
{Planck Collaboration} 2020, \mn@doi [\aap] {10.1051/0004-6361/201833910}, 641,
  A6

\bibitem[\protect\citeauthoryear{{Planck Collaboration}}{{Planck
  Collaboration}}{2021}]{Planck2018_para}
{Planck Collaboration} 2021, \mn@doi [A\&A] {10.1051/0004-6361/201833910e},
  652, C4

\bibitem[\protect\citeauthoryear{{Ratsimbazafy}, {Loubser}, {Crawford},
  {Cress}, {Bassett}, {Nichol}  \& {V{\"a}is{\"a}nen}}{{Ratsimbazafy}
  et~al.}{2017}]{Ratsimbazafy2017}
{Ratsimbazafy} A.~L.,  {Loubser} S.~I.,  {Crawford} S.~M.,  {Cress} C.~M.,
  {Bassett} B.~A.,  {Nichol} R.~C.,   {V{\"a}is{\"a}nen} P.,  2017, \mn@doi
  [MNRAS] {10.1093/mnras/stx301}, \href
  {https://ui.adsabs.harvard.edu/abs/2017MNRAS.467.3239R} {467, 3239}

\bibitem[\protect\citeauthoryear{{Renzini}}{{Renzini}}{2006}]{Renzini2006}
{Renzini} A.,  2006, \mn@doi [ARA\&A] {10.1146/annurev.astro.44.051905.092450},
  \href {https://ui.adsabs.harvard.edu/abs/2006ARA&A..44..141R} {44, 141}

\bibitem[\protect\citeauthoryear{{Riess} et~al.,}{{Riess}
  et~al.}{2022}]{Riess2022}
{Riess} A.~G.,  et~al., 2022, \mn@doi [ApJL] {10.3847/2041-8213/ac5c5b}, \href
  {https://ui.adsabs.harvard.edu/abs/2022ApJ...934L...7R} {934, L7}

\bibitem[\protect\citeauthoryear{{Said} et~al.,}{{Said}
  et~al.}{2025}]{Said2025}
{Said} K.,  et~al., 2025, \mn@doi [\mnras] {10.1093/mnras/staf700}, \href
  {https://ui.adsabs.harvard.edu/abs/2025MNRAS.539.3627S} {539, 3627}

\bibitem[\protect\citeauthoryear{{Stern}, {Jimenez}, {Verde}, {Stanford}  \&
  {Kamionkowski}}{{Stern} et~al.}{2010}]{Stern2010}
{Stern} D.,  {Jimenez} R.,  {Verde} L.,  {Stanford} S.~A.,   {Kamionkowski} M.,
   2010, \mn@doi [ApJS] {10.1088/0067-0049/188/1/280}, \href
  {https://ui.adsabs.harvard.edu/abs/2010ApJS..188..280S} {188, 280}

\bibitem[\protect\citeauthoryear{{Thomas}, {Maraston}, {Schawinski}, {Sarzi}
  \& {Silk}}{{Thomas} et~al.}{2010}]{Thomas2010}
{Thomas} D.,  {Maraston} C.,  {Schawinski} K.,  {Sarzi} M.,   {Silk} J.,  2010,
  \mn@doi [MNRAS] {10.1111/j.1365-2966.2010.16427.x}, \href
  {https://ui.adsabs.harvard.edu/abs/2010MNRAS.404.1775T} {404, 1775}

\bibitem[\protect\citeauthoryear{{Tomasetti} et~al.,}{{Tomasetti}
  et~al.}{2023}]{Tomasetti2023}
{Tomasetti} E.,  et~al., 2023, \mn@doi [A\&A] {10.1051/0004-6361/202346992},
  \href {https://ui.adsabs.harvard.edu/abs/2023A&A...679A..96T} {679, A96}

\bibitem[\protect\citeauthoryear{{Vazdekis}, {S{\'a}nchez-Bl{\'a}zquez},
  {Falc{\'o}n-Barroso}, {Cenarro}, {Beasley}, {Cardiel}, {Gorgas}  \&
  {Peletier}}{{Vazdekis} et~al.}{2010}]{Vazdekis2010}
{Vazdekis} A.,  {S{\'a}nchez-Bl{\'a}zquez} P.,  {Falc{\'o}n-Barroso} J.,
  {Cenarro} A.~J.,  {Beasley} M.~A.,  {Cardiel} N.,  {Gorgas} J.,   {Peletier}
  R.~F.,  2010, \mn@doi [MNRAS] {10.1111/j.1365-2966.2010.16407.x}, \href
  {http://saaoads.chpc.ac.za/abs/2010MNRAS.404.1639V} {404, 1639}

\bibitem[\protect\citeauthoryear{{Vazdekis} et~al.,}{{Vazdekis}
  et~al.}{2015}]{Vazdekis2015}
{Vazdekis} A.,  et~al., 2015, \mn@doi [MNRAS] {10.1093/mnras/stv151}, \href
  {http://adsabs.harvard.edu/abs/2015MNRAS.449.1177V} {449, 1177}

\bibitem[\protect\citeauthoryear{{Veale}, {Ma}, {Greene}, {Thomas},
  {Blakeslee}, {McConnell}, {Walsh}  \& {Ito}}{{Veale}
  et~al.}{2017}]{Veale2017}
{Veale} M.,  {Ma} C.-P.,  {Greene} J.~E.,  {Thomas} J.,  {Blakeslee} J.~P.,
  {McConnell} N.,  {Walsh} J.~L.,   {Ito} J.,  2017, \mn@doi [\mnras]
  {10.1093/mnras/stx1639}, \href
  {https://ui.adsabs.harvard.edu/abs/2017MNRAS.471.1428V} {471, 1428}

\bibitem[\protect\citeauthoryear{{Verde}, {Treu}  \& {Riess}}{{Verde}
  et~al.}{2019}]{Verde2019}
{Verde} L.,  {Treu} T.,   {Riess} A.~G.,  2019, \mn@doi [Nature Astronomy]
  {10.1038/s41550-019-0902-0}, \href
  {https://ui.adsabs.harvard.edu/abs/2019NatAs...3..891V} {3, 891}

\bibitem[\protect\citeauthoryear{{Wetzel}, {Tinker}  \& {Conroy}}{{Wetzel}
  et~al.}{2012}]{Wetzel2012}
{Wetzel} A.~R.,  {Tinker} J.~L.,   {Conroy} C.,  2012, \mn@doi [\mnras]
  {10.1111/j.1365-2966.2012.21188.x}, \href
  {https://ui.adsabs.harvard.edu/abs/2012MNRAS.424..232W} {424, 232}

\bibitem[\protect\citeauthoryear{{Zhou} et~al.,}{{Zhou}
  et~al.}{2023}]{Zhou2023}
{Zhou} R.,  et~al., 2023, \mn@doi [\aj] {10.3847/1538-3881/aca5fb}, \href
  {https://ui.adsabs.harvard.edu/abs/2023AJ....165...58Z} {165, 58}

\bibitem[\protect\citeauthoryear{{de Jong} et~al.,}{{de Jong}
  et~al.}{2022}]{deJong2022}
{de Jong} R.~S.,  et~al., 2022, in {Evans} C.~J.,  {Bryant} J.~J.,   {Motohara}
  K.,  eds,  Society of Photo-Optical Instrumentation Engineers (SPIE)
  Conference Series Vol. 12184, Ground-based and Airborne Instrumentation for
  Astronomy IX. p. 1218414, \mn@doi{10.1117/12.2628965}

\bibitem[\protect\citeauthoryear{{de la Rosa}, {de Carvalho}, {Vazdekis}  \&
  {Barbuy}}{{de la Rosa} et~al.}{2007}]{delaRosa2007}
{de la Rosa} I.~G.,  {de Carvalho} R.~R.,  {Vazdekis} A.,   {Barbuy} B.,  2007,
  \mn@doi [\aj] {10.1086/509502}, \href
  {https://ui.adsabs.harvard.edu/abs/2007AJ....133..330D} {133, 330}

\makeatother
\end{thebibliography}



\appendix

\section{Nominal velocity dispersion values}
\label{velocitydisp}

\begin{figure}
\centering
\includegraphics[scale=0.36, clip]{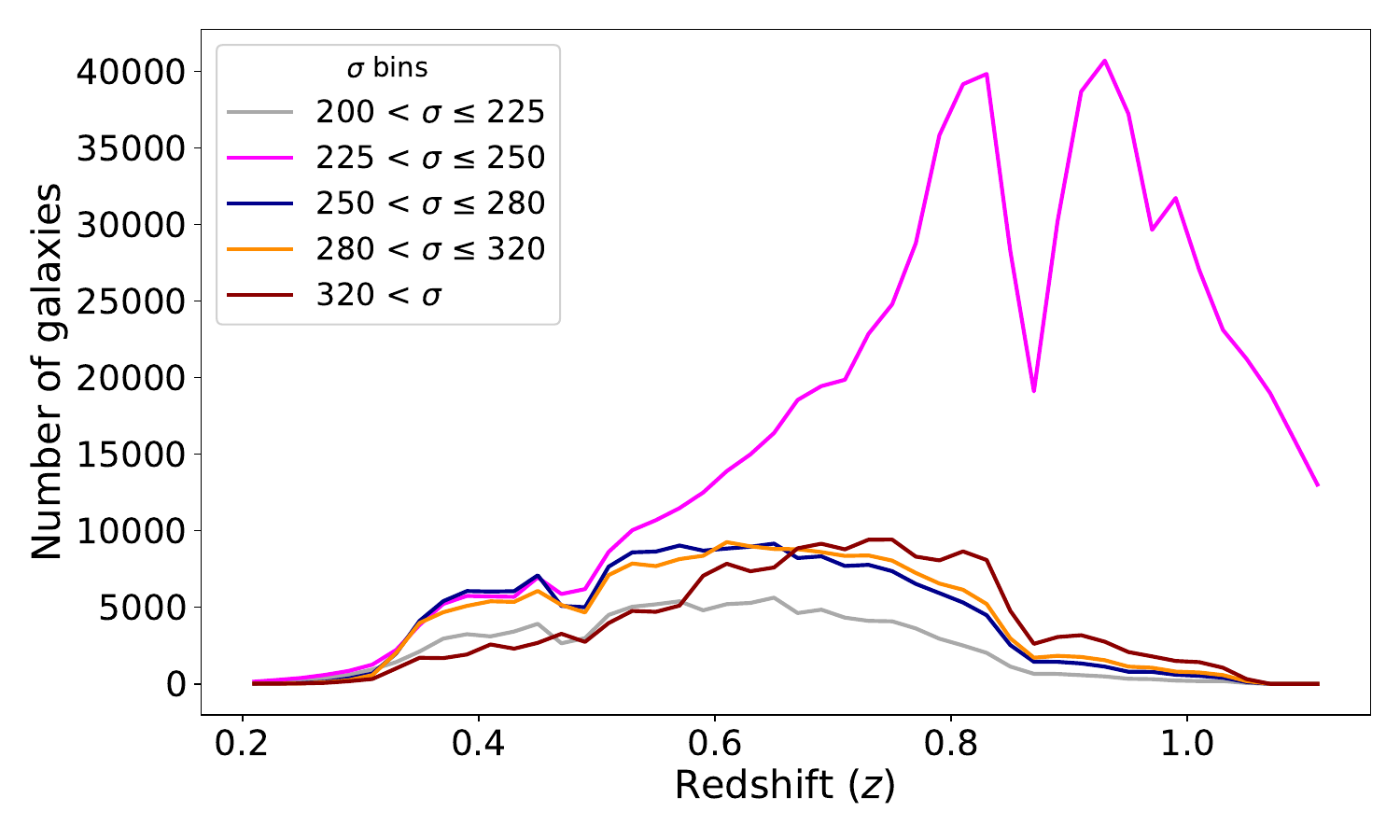}
   \caption{The selected sample divided into velocity dispersion ($\sigma$) bins (as used directly from the VAC). This is a replica of Figure \ref{fig:selection2} (bottom panel), but without the removal of the galaxies with the default velocity dispersion of 250 km s$^{-1}$.}
\label{fig:selection3}
\end{figure}

Here, we show a version of Figure \ref{fig:selection2} (bottom panel), but without removing galaxies with the nominal/default velocity dispersion of 250 km s$^{-1}$. These are the objects for which the velocity dispersion could not be accurately fitted. Figure \ref{fig:selection3} shows how this will affect the velocity dispersion distribution in the 240 $< z \leq 250$ km s$^{-1}$ bin, therefore, we use this bin as 240 $< z < 250$ km s$^{-1}$ throughout the paper.

\section{Aperture effects}
\label{apertures}

In this section, we test the effect of aperture fill (or cover) on our results. DESI’s 1.5$\arcsec$ diameter fibres, assuming standard cosmology, correspond to $\sim$6.7 kpc (radius 3.35 kpc) at $z=0.3$ -- covering at least the central bulge region of a massive galaxy. At $z=1.0$, it corresponds to $\sim$12 kpc (radius 6 kpc) -- often most or all of the light of the galaxy. This difference in coverage becomes particularly problematic when measuring properties with a strong gradient across the galaxy. For the D4000$_{\rm n}$ measurements, plotted as a function of $z$, we need to test whether there might be an artificial trend with $z$ even if the true population of the galaxy did not evolve, i.e., purely as a result of the fibre sampling a larger fraction of the galaxy at higher $z$. Theoretically, for a massive early-type galaxy, we can consider a de Vaucouleurs surface brightness profile, a mild negative radial gradient in D4000$_{\rm n}$, and a shallow velocity dispersion decline with radius, then compute the luminosity-weighted values inside the fibre radius as a function of $z$ (using a simple kpc/arcsec approximation). We can also use Integral Field Unit (IFU) observations of nearby massive, passive galaxies to quantitatively test the effect. 

\begin{figure}
\centering
\includegraphics[scale=0.46, clip]{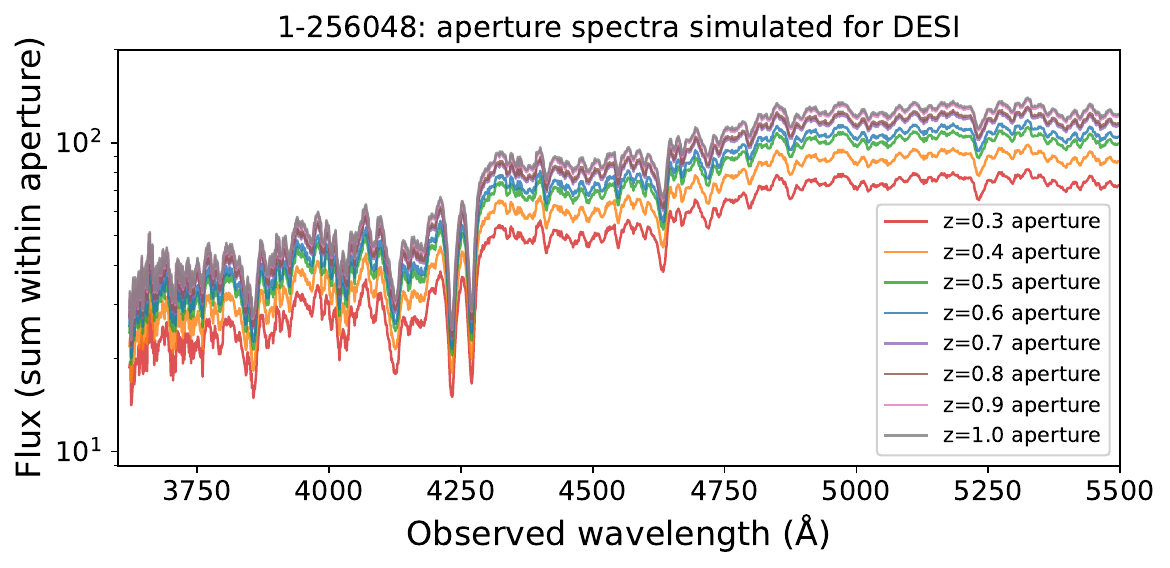} \\
\includegraphics[scale=0.5, clip]{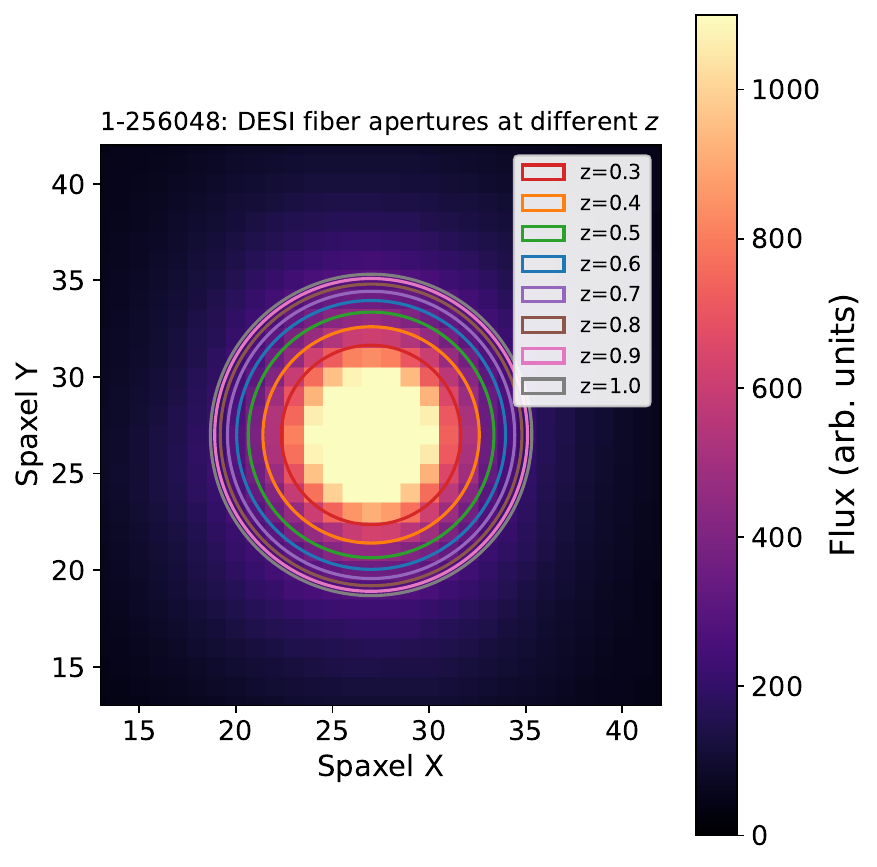}
   \caption{We extract integrated 1D spectra from simulated DESI-sized apertures at different redshifts from $z=$ 0.3 to 1.0 from a MaNGA galaxy with properties typical of our selected sample. From this, we measure the change in light-weighted D4000$_{\rm n}$ measurements as a result of different fibre cover. The extracted, integrated 1D spectra are shown in the top panel, and the apertures indicated on a 2D flux map in the bottom panel.}
\label{fig:Aperture_spectra}
\end{figure}	

We used the Sloan Digital Sky Survey IV Mapping Nearby Galaxies at the APO (MaNGA) survey \citep{Bundy2015} \texttt{Firefly} Value-Added Catalogue\footnote{\url{https://www.sdss4.org/dr16/manga/manga-data/manga-firefly-value-added-catalog/}}, containing measurements for 4675 MaNGA galaxies from DR16. The catalogue contains basic galaxy information (e.g., galaxy ID, galaxy mass), global parameters (e.g., central age), gradient parameters (e.g., age gradient) and spatially resolved quantities (e.g., 2D age maps) by \citet{Goddard2017} and \citet{Parikh2018}. From this VAC, we select nearby ($z < 0.1$) galaxies with stellar mass $\log M_{\star}/M_{\sun} > 11$ and with light-weighted stellar population age > 8 Gyr. We then extract the MaNGA DR16 flux cubes for the 47 galaxies that match the criteria using the \texttt{Marvin} Webservice\footnote{\url{https://magrathea.sdss.org/marvin/}}. We then use the \texttt{Marvin} Python tools to simulate DESI-sized apertures at different redshifts from $z=$ 0.3 to 1.0. From these apertures, we extract integrated 1D spectra, and measure the light-weighted D4000$_{\rm n}$ to analyse the effect of different fibre coverage. 

We show a typical example in Figure \ref{fig:Aperture_spectra}. The extracted, integrated 1D spectra are shown in the top panel, and the apertures are indicated on a 2D flux map in the bottom panel. Because the measurements are light-weighted properties, with surface brightness declining steeply towards the outskirts of the galaxy, we find that the typical decrease in measured D4000$_{\rm n}$ for the integrated aperture-extracted spectra across $z=$ 0.3 to 1.0 is only 1 -- 2\%. We find similar results for velocity dispersion. In addition, note that DESI’s $\sim$1.1$\arcsec$ seeing can smear light into the fibre, slightly increasing effective aperture, especially at lower $z$. Combined, we can conclude that the effect of fibre cover is, therefore, negligible on our results. 

\section{Broadening correction factor for D4000$_{\rm n}$}
\label{broadening}

We also derive broadening functions to measure the effect of velocity dispersion (broadening) on the D4000$_{\rm n}$ measurements, even if it is expected to be subtle. We follow a method similar to \citet{delaRosa2007, Clerici2024}, and use model spectra to derive the combined broadening function. We created a library of high-resolution SSP model spectra from MILES \citep{Vazdekis2010}, which has high intrinsic resolution (low $\sigma$). We used a grid of spectra from SSP ages $\sim$8 to 14 Gyr, and metallicities $[$Fe/H$]$ of --0.4, 0.0, and +0.22. We then apply artificial velocity dispersion broadening by convolving with Gaussians from $\sigma$ = 0 to 400 km s$^{-1}$ in 10 km s$^{-1}$ steps. The instrumental resolution ($\lambda/\Delta \lambda$) of DESI varies as a function of wavelength from 2000 in the blue arm to approximately 5000 in the near-infrared arm. It is comparable with the resolution of the models, and the broadening from instrumental resolution is negligible compared to that from velocity dispersion, particularly for the typical velocity dispersions of our massive, passive galaxies.

\begin{figure}
\centering
\includegraphics[scale=0.40, clip]{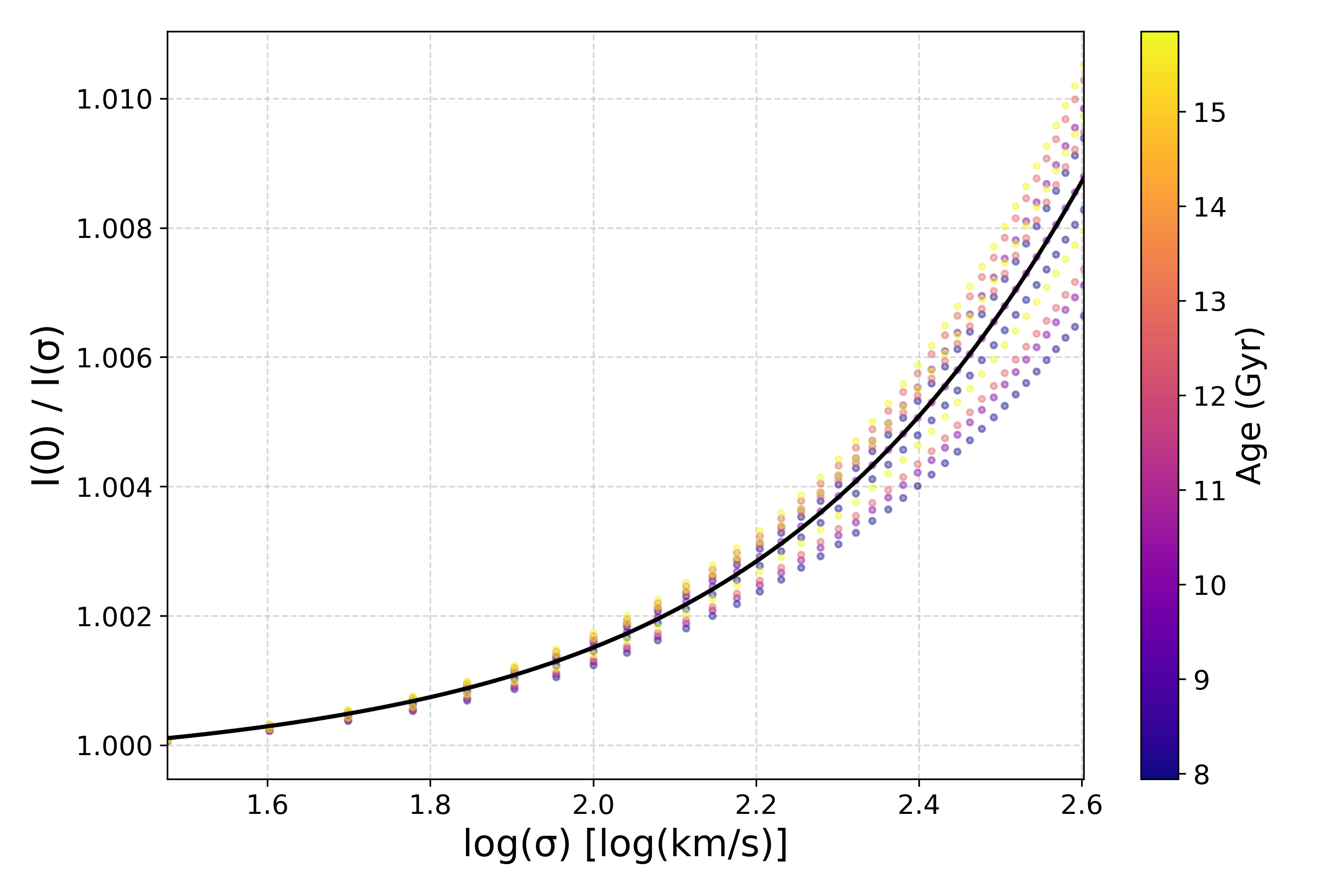} 
   \caption{We calculate the ratio between the value of the index in the non-broadened ($\sigma=0)$ versus the broadened template ($\sigma$) for templates with ages above $\sim$8 Gyr, and at the three different metallicities (thus there are three different lines for every colour that represents the age). We then also fit a combined broadening correction (using a 4-rth order polynomial) shown with the solid line.}
\label{fig:Broadening_spectra}
\end{figure}

From the artificially broadened spectra, we measure the D4000$_{\rm n}$ index and fit a 4-rth order polynomial to the broadening correction, i.e., the ratio between the index value in the non-broadened spectrum ($\sigma$=0) versus the measurement from the broadened template ($\sigma$). We show this in Figure \ref{fig:Broadening_spectra} for SSP templates with ages greater than $\sim$ 8 Gyr, and for the three different metallicities (thus, there are three different lines for every colour that represents the age). We then also fit a combined broadening correction (using a 4-rth order polynomial) shown with the solid line. This combined broadening function is given by $a_{4} x^{4} + a_{3} x^{3} + a_{2} x^{2} + a_{1} x + a_{0}$, where $a_{4} = 0.0028$, $a_{3} = -0.0164$, $a_{2} = -0.0374$, $a_{1} = -0.0384$, and $a_{0} = 1.0145$, and where $x=\log(\sigma)$. The figure shows that even at high velocity dispersion (> 280 km s$^{-1}$) there is a negligible correction of $<1\%$ for D4000$_{\rm n}$ measurements. Broadening, therefore, has no effect on the derived D4000$_{\rm n}$ -- $z$ relation.

\section{Homogeneous CC OHD using D4000$_{\rm n}$}
\label{OHDD4000}

In the full-spectrum fitting approach used for CC studies, it can be more difficult to accurately separate the various contributions to the systematic errors from the choices made in full-spectrum fitting or the degeneracies between them. In addition, D4000$_{\rm n}$ generally offers tighter constraints than full-spectrum fitting (Figure \ref{fig:Hz}). For convenience, we summarise all measurements made using D4000$_{\rm n}$ that can be used as a more homogeneous set of CC OHD datapoints to test cosmological models in Table \ref{tableOHDD4000}. The data are from this paper, and \citet{Moresco2012, Moresco2015, Moresco2016}, as well as \citet{Loubser2025}, although it should be noted that the latter only uses BCGs. 

\begin{table}
\caption{CC measurements made using D4000$_{\rm n}$. The measurement indicated by $^{*}$ was derived using BCGs only.}    
\label{tableOHDD4000}      
\centering                         
\begin{tabular}{c c c} 
\hline
Reference & $z$ & $H(z)$  \\
 &  &  $\rm\ km\ s^{-1}\ Mpc^{-1}$ \\
\hline 
\citet{Moresco2012} & 0.179 & 75 $\pm$ 4 \\
\citet{Moresco2012} & 0.199 & 75 $\pm$ 5 \\
\citet{Moresco2012} & 0.352 & 83 $\pm$ 14 \\
\citet{Moresco2016} & 0.380 & 83 $\pm$ 13.5 \\
\citet{Moresco2016} & 0.400 & 77 $\pm$ 10.2 \\
\citet{Moresco2016} & 0.425 & 87.1 $\pm$ 11.2 \\
\citet{Moresco2016} & 0.445 & 92.8 $\pm$ 12.9 \\
This paper & 0.46 &  $ 88.48 \pm\ 0.57(\rm stat) \pm 12.32(\rm syst)$   \\
\citet{Moresco2016} & 0.478 & 80.9 $\pm$ 9 \\
\citet{Moresco2012} & 0.593 & 104 $\pm$ 13 \\
This paper & 0.67 &  $ 119.45 \pm\ 6.39(\rm stat) \pm 16.64(\rm syst)$   \\
\citet{Moresco2012} & 0.680 & 92 $\pm$ 8 \\
\citet{Moresco2012} & 0.781 & 105 $\pm$ 12 \\
This paper & 0.83 &  $108.28 \pm\ 10.07(\rm stat) \pm 15.08(\rm syst)$ \\
\citet{Moresco2012} & 0.875 & 125 $\pm$ 17 \\
\citet{Moresco2012} & 1.037 & 154 $\pm$ 20 \\
\citet{Moresco2015} & 1.363 & 160 $\pm$ 33.6 \\
\citet{Moresco2015} & 1.965 & 186.5 $\pm$ 50.4 \\
\hline
\citet{Loubser2025} & 0.500 & $72.1 \pm\ 33.9(\rm stat) \pm 7.3(\rm syst)^{*}$ \\
\hline
\end{tabular}
\end{table}	


\bsp 
\label{lastpage}
\end{document}